\pdfoutput=1
\documentclass[12pt]{article}
\usepackage[margin=1in]{geometry}
\usepackage[utf8]{inputenc}
\usepackage[T1]{fontenc}
\usepackage{amsmath,amssymb,amsthm,mathtools,bm}
\mathtoolsset{showonlyrefs=true}
\usepackage{graphicx}
\usepackage{booktabs,multirow,array}
\usepackage[labelfont=bf,labelsep=period]{caption}
\makeatletter
\renewcommand\subsection{\@startsection{subsection}{2}{\z@}%
  {-3.25ex\@plus -1ex \@minus -.2ex}{1.5ex \@plus .2ex}{\normalfont\large\bfseries\raggedright}}
\renewcommand\paragraph{\@startsection{paragraph}{4}{\z@}%
  {1.2ex \@plus .4ex \@minus .2ex}{-1em}{\normalfont\normalsize\bfseries}}
\makeatother
\usepackage{setspace}
\usepackage{xcolor}
\usepackage{tikz}
\usetikzlibrary{arrows.meta,positioning}
\usepackage{algorithm}
\usepackage{algpseudocode}
\usepackage[natbibapa]{apacite}
\makeatletter
\let\APAC@orig@bibcite\bibcite
\def\bibcite#1#2{\APAC@orig@bibcite{#1}{#2}\shortcites{#1}}
\makeatother
\usepackage[hidelinks]{hyperref}

\let\hat\widehat
\let\tilde\widetilde
\newcommand{\argmin}{\operatornamewithlimits{arg\,min}}
\newcommand{\diag}{\operatorname{diag}}
\newcommand{\tr}{\operatorname{tr}}
\newcommand{\Cov}{\operatorname{Cov}}

\newcommand{\CPDAG}{\operatorname{CPDAG}}

\newcommand{\I}{\mathbf I}
\newcommand{\bzero}{\mathbf 0}
\newcommand{\bone}{\mathbf 1}
\newcommand{\Y}{\mathbf Y}
\newcommand{\y}{\mathbf y}
\newcommand{\bS}{\mathbf S}
\newcommand{\bZ}{\mathbf Z}
\newcommand{\bfeta}{\bm\eta}
\newcommand{\bfzeta}{\bm\zeta}
\newcommand{\bfeps}{\bm\varepsilon}
\newcommand{\bftheta}{\bm\theta}
\newcommand{\calG}{\mathcal G}
\newcommand{\calC}{\mathcal C}
\newcommand{\calD}{\mathcal D}
\newcommand{\calM}{\mathcal M}
\newcommand{\calN}{\mathcal N}

\newcommand{\calE}{\mathcal E}

\newcommand{\calV}{\mathcal V}

\newcommand{\R}{\mathbb R}

\theoremstyle{plain}
\newtheorem{theorem}{Theorem}

\newtheorem{corollary}{Corollary}
\theoremstyle{definition}
\newtheorem{definition}{Definition}
\newtheorem{assumption}{Assumption}
\newtheorem{example}{Example}
\theoremstyle{remark}
\newtheorem{remark}{Remark}

\title{\Large Data-Driven Learning of Directed Acyclic Graphs in Gaussian Linear Structural Equation Models with Latent Variables}
\date{}

\author{%
  {\normalsize Siliang Zhang\textsuperscript{1},
  Yunxiao Chen\textsuperscript{2}, and Irini Moustaki\textsuperscript{2}}\\[0.6em]
  {\small\textsuperscript{1}School of Statistics, East China Normal University}\\
  {\small\textsuperscript{2}Department of Statistics, London School of Economics and Political Science}}
\hypersetup{pdftitle={Data-Driven Learning of Directed Acyclic Graphs in Gaussian Linear Structural Equation Models with Latent Variables},pdfauthor={Siliang Zhang, Yunxiao Chen, Irini Moustaki}}

\begin{document}
\maketitle

\begin{abstract}
\noindent 
Structural equation modeling is typically performed in a confirmatory manner, requiring a priori specification of both the measurement model and the path diagram connecting the structural variables. In this paper, we study a semi-exploratory setting within a general Gaussian linear structural equation model with latent variables, where the loading structure in the measurement model is known, but the structural path diagram follows an unknown directed acyclic graph (DAG) to be learned from data. First, we establish the identifiability of the structural model, showing that the unknown DAG is identifiable only up to its Markov equivalence class. We then propose DAGGLE (\underline{D}irected \underline{A}cyclic \underline{G}raph learning in \underline{G}aussian \underline{L}inear structural \underline{E}quation models), a two-stage likelihood-based approach for simultaneous DAG structure learning and parameter estimation. In the first stage, we screen candidate edges via continuous optimization of a penalized marginal likelihood. In the second stage, we further prune the remaining edges using a BIC-guided backward search over the space of Markov equivalence classes. Under mild regularity and search conditions, DAGGLE recovers the true Markov equivalence class with probability tending to one.
Finally, we demonstrate the effectiveness of our approach through extensive numerical simulations and an application to the European Social Survey (Trust in Justice module), where DAGGLE yields a model that aligns with substantive theory while improving overall fit.

\end{abstract}


\section{Introduction}\label{sec:intro}

Structural equation modeling is concerned with estimating (linear and non-linear) relationships between constructs (latent variables/factors) which are measured through imperfect observed indicators (items). Examples of constructs are cognitive abilities, personality traits, institutional trust, political attitudes, social values, etc. Structural equation models (SEMs) consist of a measurement model that links the latent variables to their observed indicators and a structural model that formalizes the hypothesized relationships among the constructs \citep{bollen1989structural, bartholomew2011latent}.  

Often the measurement model is a confirmatory factor analysis model \citep{joreskog:69} which postulates specific relationships between the latent variables and the indicators, assuming a prespecified pattern for the model parameters (factor loadings, error variances). It is common practice to aim to improve the fit of the measurement model by freeing some of the constraints and estimating more of the model parameters, leading to an interplay between exploratory and confirmatory factor analysis. Traditionally, the model fit improvement is guided by modification indices \citep{MacCallum1986,Sorbom1989}, which are score tests that quantify how much the chi-square test statistic would drop when a parameter is freed to be estimated. \citet{MacCallum1986} states that even under favorable conditions (hypothesized model close to the true model, large sample size), models arising from specification searches must be viewed with caution. Less often, the structural model is modified based on data evidence. 

In many applications, validated instruments provide a defensible measurement model, whereas the structural paths remain uncertain. Researchers may then compare a few prespecified models or modify a fitted model using modification indices, similarly to measurement model improvement. This motivates learning the structural paths while retaining the measurement structure and accounting for measurement error.
Much of the literature on exploratory SEM focuses instead on improving the fit of the measurement model to prevent misspecification from biasing the structural parameter estimates \citep{AsparouhovMuthen2009, Morinetal:13, MarshMorinParkerKaur2014}.

Directed acyclic graphs (DAGs) are commonly used to model directional relationships in structural models, which form the foundation of recursive SEMs \citep[Chapter~4]{bollen1989structural}. In this paper, we consider a semi-exploratory framework in which the measurement structure is prespecified, and the measurement model is identified, while the structural DAG is learned subject to any prior restrictions. Estimation uses the marginal likelihood of the observed indicators, without treating estimated factor scores as observed variables. The framework distinguishes path directions in the structural model identified from the observed indicator distribution from those supplied by theory-driven restrictions. The European Social Survey application illustrates how this distinction informs the assessment of the structural equation model  specified for these data in \citet{jackson2013policing}, following the methodological-substantive synergy advocated by \citet{MarshHau2007}. The proposed methodology aims to provide an easy-to-use exploratory tool to allow for such a synergy to be carried out under a sound statistical framework. 

Our contribution is threefold. First, we formalize identification of the structural DAG under an identified Gaussian linear SEM measurement model. The result allows prespecified cross-loadings and correlated measurement errors. Under a faithfulness condition on the latent distribution stated in Section~\ref{subsec:identifiability}, the DAG is identified up to its Markov equivalence class, which can be further narrowed by prior structural restrictions. Second, we develop DAGGLE (Directed Acyclic Graph learning in Gaussian Linear structural Equation models), which adapts the screening-and-pruning strategy of \citet{ZhangChenMoustaki2026} to the exact Gaussian marginal likelihood. DAGGLE jointly estimates measurement and structural parameters through continuous screening with a truncated $\ell_1$ penalty \citep{shen2012likelihood} and a smooth acyclicity constraint \citep{ZhengEtAl2018}, followed by BIC-guided backward pruning over admissible equivalence classes. Its consistency follows by adapting the argument of \citet{ZhangChenMoustaki2026} under the conditions discussed in Section~\ref{subsec:consistency}. Third, we analyze how factor-score substitution and measurement error affect graph learning, benchmark our procedure against two-step alternatives in simulations, and apply it to the Trust in Justice module of the European Social Survey (Round 5).

Our work directly connects with the vast literature on causal discovery and DAG learning for fully observed variables, also known as Bayesian networks. This field has traditionally relied on constraint-based methods such as the PC algorithm, score-based methods such as greedy equivalence search (GES), or hybrid combinations \citep{SpirtesGlymourScheines2000, chickering2002optimal, pearl2009causality}. More recently, \cite{ZhengEtAl2018} recast DAG learning as a continuous optimization problem, which tends to outperform traditional methods based on combinatorial searches. For latent variables, MIMBuild \citep{SilvaScheinesGlymourSpirtes2006} uses a pure measurement model and learns the latent structure by constraint-based search or GES guided by a BIC score approximated using Structural EM. Our framework instead accommodates identified measurement models with prespecified cross-loadings and correlated measurement errors, and combines continuous screening with BIC-guided backward pruning based on the exact Gaussian marginal likelihood under prior structural restrictions. Our identifiability theory provides a direct analog to the fully observed case \citep[see, e.g.,][]{pearl2009causality}. 
 
This work is also closely related to network psychometrics. The network approach to psychology typically utilizes Gaussian graphical models or Ising models to characterize conditional dependencies among observed elements, such as symptoms or attitudes \citep{BorsboomCramer2013, EpskampBorsboomFried2018, MarsmanEtAl2018, BorsboomEtAl2021, MarsmanRhemtulla2022, MarsmanHuthWaldorpNtzoufras2022, Borsboom2022, MarsmanVanDenBerghHaslbeck2025}. These network models have been combined with latent variables in unified SEMs \citep{EpskampRhemtullaBorsboom2017} and compared with latent variable models \citep{VanBorkEtAl2021, WaldorpMarsman2022}, but this literature is largely confined to undirected conditional dependence. Directed graph learning, including cyclic causal discovery, has also been studied for psychological data \citep{ParkWaldorpRyan2024}. By targeting a directed acyclic structure at the latent level while explicitly accounting for measurement error, our approach provides a new exploratory and computationally competitive tool for modeling directional pathways among higher-level latent variables. Such a directed structure encodes a different family of conditional independence relations from an undirected network, as discussed in Section~\ref{subsec:mec}.

The current framework assumes a linear factor model for the measurement component, which requires the observed variables to be continuous. Separately, a growing body of work addresses causal discovery and graphical modeling for categorical survey data \citep[e.g.,][]{LuoMoffaKuipers2021, Castelletti2024, ni2025causal}. However, these approaches do not incorporate explicit measurement models and thus cannot accommodate latent variables measured by categorical indicators. Directed structures among discrete latent variables have also been studied in cognitive diagnosis and causal representation learning \citep*{GuXu2019,LeeGu2024,zhang2026discrete}. More recently, \citet{ZhangChenMoustaki2026} develop a general framework that combines penalized continuous screening with backward pruning over Markov equivalence classes, tailored to categorical indicators via composite likelihood. The present paper carries this strategy into the Gaussian linear SEM and develops it for structural equation modeling and network psychometrics. Unlike \citet{ZhangChenMoustaki2026}, who assume that each indicator measures a single latent variable, our model and theory allow identified measurement models with prespecified cross-loadings and correlated errors. Furthermore, we make explicit how prior structural restrictions inform the identified graph and the consistency of the backward search. 

Our work is also related to data-driven selection of paths within SEMs. Regularized SEM selects paths through sparsity-inducing penalties \citep{JacobucciGrimmMcArdle2016,HuangChenWeng2017}, and knockoff methods select covariates in latent regression models \citep{XieChenVonDavierWeng2024}. Learning such a sparse latent regression or multiple-indicator multiple-cause (MIMIC) model is a special case of structure learning in which every candidate path points from an observed covariate to a latent variable. Exploratory SEM relaxes the restrictions of the measurement model \citep{AsparouhovMuthen2009}, and Bayesian SEM replaces zero restrictions with informative small-variance priors \citep{MuthenAsparouhov2012}. In these approaches, the direction of each candidate path is specified in advance, whereas the present paper learns the directions among the latent variables up to their Markov equivalence class. Exploratory SEM and the present framework relax complementary components of a SEM. Exploratory SEM retains a confirmatory structural model while estimating rotated loadings across all indicators, a strategy widely adopted because restricting each indicator to a single loading often yields poor fit \citep{MarshMorinParkerKaur2014}. In contrast, the present framework keeps the measurement model confirmatory and learns the structural model.
 
The proposed truncated $\ell_1$ penalty of the screening stage approximates the number of nonzero coefficients. Unlike the $\ell_1$ penalty, it does not shrink coefficients above a truncation level, and its global minimizer selects the nonzero coefficients consistently \citep{shen2012likelihood,shen2013constrained}. Penalties of this kind have also been used to learn DAGs over observed variables, including the $\ell_0$ penalty, concave penalties, and truncated $\ell_1$ constraints \citep{vandegeer2013l0,aragam2015concave,yuan2019constrained}. In psychometrics, the truncated $\ell_1$ penalty has been used to detect differential item functioning across multiple groups, because it avoids the heavy penalization of large group differences under the $\ell_1$ penalty \citep{LyuWangXu2025}.

The remainder of the paper is organized as follows. Section~\ref{sec:model} introduces the model, reviews Markov equivalence, and defines the graph learning problem under background constraints. Section~\ref{sec:alg} develops the two-stage DAGGLE algorithm and compares it with alternative methods. Section~\ref{sec:iden} establishes the identifiability of the structural DAG and discusses the asymptotic consistency of DAGGLE. Sections~\ref{sec:simulation} and \ref{sec:real} report the simulation study and the application to the European Social Survey, and Section~\ref{sec:dis} concludes with a discussion of methodological extensions. Technical proofs and auxiliary analyses are given in the Supplementary Material S1--S4, which is available from the authors upon request. A Python package and replication materials will be made publicly available upon acceptance of the manuscript.

\section{Problem Setting}\label{sec:model}

We consider a Gaussian linear structural equation model for the latent variables $\bfeta=(\eta_1,\ldots,\eta_m)^\top$, measured by centered observed indicators $\Y=(Y_1,\ldots,Y_p)^\top$. For ease of exposition, we treat all structural variables as latent. Section~S2 extends the model to observed exogenous covariates. We study a semi-exploratory setting in which the measurement structure is prespecified, while the latent variables follow an unknown directed acyclic graph (DAG). In what follows, we present the model formulation, discuss its observational equivalence limits, incorporate prior structural knowledge, and state the DAG learning problem.

\subsection{Model}\label{subsec:model}\label{subsec:measurement}\label{subsec:structural}

\paragraph{Measurement model.}
The measurement model follows a linear factor model,
\begin{equation}\label{eq:measurement}
  \Y=\Lambda\bfeta+\bfeps,
  \qquad
  \bfeps\sim\calN(\bzero,\Psi),
\end{equation}
where $\Lambda \in \mathbb{R}^{p\times m}$ is the factor loading matrix and the measurement error $\boldsymbol{\varepsilon} = (\varepsilon_1,\ldots,\varepsilon_p)^\top$ is independent of $\boldsymbol{\eta}$. We assume the latent variables follow $\boldsymbol{\eta} \sim \mathcal{N}(\mathbf{0},\Phi)$ with a positive definite covariance matrix $\Phi$. The model-implied covariance matrix of the observed indicators is therefore $\Sigma(\bftheta)=\Lambda\Phi\Lambda^\top+\Psi$, where $\bftheta$ collects the model parameters specified below.

The measurement structure is prespecified. A binary configuration matrix $Q\in\{0,1\}^{p\times m}$ encodes known structural zeros in the loading matrix, where $\Lambda_{jk}=0$ whenever $Q_{jk}=0$. The matrix $\Psi$ need not be diagonal, thereby accommodating correlated errors often encountered in confirmatory factor analysis. To resolve scale and sign indeterminacies, we set $\diag(\Phi)=\bone_m$ and require one prespecified nonzero entry in each column of $\Lambda$ to be positive. Section~\ref{subsec:identifiability} states the condition under which these restrictions identify the measurement model.

\paragraph{Structural model.}
We specify the directional relationships among the latent variables $\bfeta$ through linear structural equations,
\begin{equation}\label{eq:structural}
  \bfeta=B^\top\bfeta+\bfzeta,
  \qquad
  \bfzeta\sim\calN(\bzero,\Omega),
\end{equation}
where $B = (\beta_{ij})_{m\times m}$ is a matrix of path coefficients with zero diagonal entries, and $\boldsymbol{\zeta} = (\zeta_1,\ldots,\zeta_m)^\top$ is a vector of mutually independent disturbances with diagonal covariance matrix $\Omega = \mathrm{diag}(\omega_1,\ldots,\omega_m)$ ($\omega_j > 0$). The entry $\beta_{ij}$ denotes the direct effect of $\eta_i$ on $\eta_j$, while the zero diagonal ensures that no variable exerts a direct effect on itself. Let $\bftheta=(\Lambda,\Psi,B,\Omega)$ collect the measurement parameters $(\Lambda,\Psi)$ and structural parameters $(B,\Omega)$, subject to the stated restrictions and normalizations.

The structural model \eqref{eq:structural} differs from a classical confirmatory SEM in two respects. First, a confirmatory SEM prespecifies zero constraints on the off-diagonal entries of $B$ from a path diagram, with $\beta_{ij}=0$ if there is no directed edge from variable $i$ to variable $j$. In the current setting, the path diagram is unknown, and these zero constraints are learned from data. Second, classical SEMs divide the structural variables into exogenous and endogenous variables according to their roles in the path diagram. We do not make this distinction during model specification. Once the path diagram is learned, the classification can be made to the extent that its edge directions are identified (Section~\ref{subsec:mec}).

The path diagram can be viewed as a directed graph, with one vertex for each structural variable and one directed edge for each arrow. A DAG is a directed graph with no directed cycles, so no variable can be reached from itself by following the arrows. Many path diagrams in practice follow a DAG, which represents hypotheses about causal relationships among the structural variables. With independent disturbances, such models are known as recursive models in the SEM literature \citep[Chapter~4]{bollen1989structural}. We focus on this setting, as formalized below.

\begin{definition}[Directed acyclic graph]\label{def:dag}
  A directed graph $\calG=(\calV,\calE)$ consists of a vertex set $\calV=\{1,\ldots,m\}$ and a set of directed edges $\calE$. An ordered pair $(i,j)\in\calE$ represents an edge directed from $i$ to $j$, denoted by $i\to j$, in which case $i$ is called a parent of $j$. A directed cycle is a sequence of edges $i\to\cdots\to i$ that returns to its origin. The graph $\calG$ is a directed acyclic graph (DAG) if it contains no directed cycle.
\end{definition}

Each path coefficient matrix $B$ induces a directed graph $\calG(B)$ on $\calV$ with edge set $\{(i,j):\beta_{ij}\neq0\}$. Let $\calD_m$ denote the set of all DAGs on $\calV$. Whenever $\calG(B)\in\calD_m$, the matrix $\I-B^\top$ is invertible, and the latent covariance matrix is given by
\begin{equation}\label{eq:latent_cov}
  \Phi(B,\Omega)=(\I-B^\top)^{-1}\Omega(\I-B)^{-1}.
\end{equation}
Thus, the indicator covariance matrix is $\Sigma(\bftheta)=\Lambda\Phi(B,\Omega)\Lambda^\top+\Psi$.

\subsection{Markov Equivalence}\label{subsec:mec}

Observational data do not always determine the direction of an edge. Distinct path diagrams can imply the same set of covariance matrices for the indicators, a phenomenon known in the SEM literature as equivalent models \citep{Stelzl1986,LeeHershberger1990,maccallum1993problem}. We describe equivalence between DAGs through the conditional independence relations they imply among the latent variables \citep{verma1990equivalence,pearl2009causality}. Two latent variables are conditionally independent given a set of other latent variables if their joint distribution given that set factorizes into the product of the two conditional distributions. Under the Gaussian structural model \eqref{eq:structural}, this holds if and only if the corresponding partial correlation is zero.

\begin{definition}[Markov equivalence class]\label{def:mec}
Two DAGs on $\mathcal{V}$ are Markov equivalent if they imply the
same conditional independence relations among the latent variables.
A graph implies such a relation if it holds under \eqref{eq:structural} for every parameter value with $\beta_{ij}=0$ whenever the graph contains no edge $i\to j$.
The Markov equivalence class (MEC) of a DAG $\mathcal{G}$, denoted by
$\mathcal{M}(\mathcal{G})$, is the set of all DAGs Markov equivalent
to $\mathcal{G}$.
\end{definition}

Under the Gaussian linear model of Section~\ref{subsec:model}, Markov-equivalent DAGs can produce the same latent covariance matrices $\Phi(B,\Omega)$ as their parameters vary, and hence the same distributions of the indicators, so the likelihood cannot distinguish them. Two vertices are adjacent if they are joined by an edge, regardless of its direction. Markov-equivalent DAGs have the same adjacencies \citep{verma1990equivalence}. An MEC $\calM(\calG)$ is uniquely represented by its completed partially directed acyclic graph (CPDAG), denoted by $\CPDAG(\calG)$. An edge of the CPDAG is directed if all members of $\calM(\calG)$ orient it in the same way. Otherwise, the edge is reversible and is left undirected \citep{andersson1997characterization,chickering2002equivalence,meek1995causal}.

\begin{figure}[t]
  \centering
  \begin{tikzpicture}[>=Stealth, thick,
      v/.style={circle, draw, minimum size=9mm, inner sep=1pt},
      el/.style={font=\small, inner sep=1.5pt}]
    \begin{scope}
      \node[v] (a1) at (0,0)    {$\eta_1$};
      \node[v] (a2) at (1,1.15) {$\eta_2$};
      \node[v] (a3) at (2,0)    {$\eta_3$};
      \draw[->] (a1) -- node[el,above left]  {$0.6$} (a2);
      \draw[->] (a2) -- node[el,above right] {$0.5$} (a3);
      \node at (1,-0.85) {(a) chain};
      \node[font=\small] at (1,-1.35) {$\omega=(1,\,0.64,\,0.75)$};
    \end{scope}
    \begin{scope}[xshift=4.4cm]
      \node[v] (b1) at (0,0)    {$\eta_1$};
      \node[v] (b2) at (1,1.15) {$\eta_2$};
      \node[v] (b3) at (2,0)    {$\eta_3$};
      \draw[->] (b2) -- node[el,above left]  {$0.6$} (b1);
      \draw[->] (b3) -- node[el,above right] {$0.5$} (b2);
      \node at (1,-0.85) {(b) reverse chain};
      \node[font=\small] at (1,-1.35) {$\omega=(0.64,\,0.75,\,1)$};
    \end{scope}
    \begin{scope}[xshift=8.8cm]
      \node[v] (c1) at (0,0)    {$\eta_1$};
      \node[v] (c2) at (1,1.15) {$\eta_2$};
      \node[v] (c3) at (2,0)    {$\eta_3$};
      \draw[->] (c2) -- node[el,above left]  {$0.6$} (c1);
      \draw[->] (c2) -- node[el,above right] {$0.5$} (c3);
      \node at (1,-0.85) {(c) fork};
      \node[font=\small] at (1,-1.35) {$\omega=(0.64,\,1,\,0.75)$};
    \end{scope}
  \end{tikzpicture}
  \caption{Three Markov-equivalent DAGs over $\eta_1,\eta_2,\eta_3$ with the adjacencies $\eta_1-\eta_2$ and $\eta_2-\eta_3$ and no collider at $\eta_2$. Edge labels denote path coefficients, and $\omega=(\omega_1,\omega_2,\omega_3)$ denotes the disturbance variances $\Omega=\diag(\omega_1,\omega_2,\omega_3)$. All three graphs imply the latent covariance matrix \eqref{eq:mec_cov} and generate identical indicator distributions.}
  \label{fig:mec_example}
\end{figure}
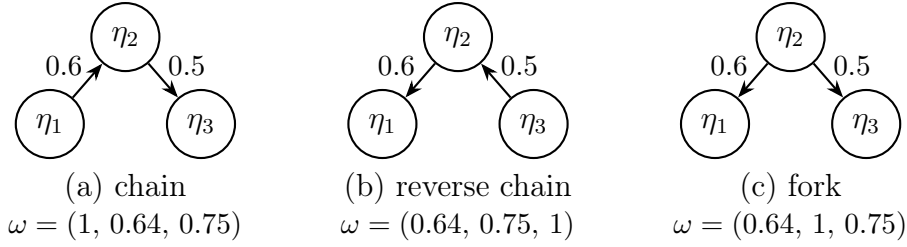

To illustrate, consider three latent variables with covariance matrix
\begin{equation}\label{eq:mec_cov}
  \Phi=\begin{pmatrix}1&0.6&0.3\\0.6&1&0.5\\0.3&0.5&1\end{pmatrix}.
\end{equation}
Because $\Phi_{13}=\Phi_{12}\Phi_{23}=0.6\times0.5=0.3$, the partial correlation between $\eta_1$ and $\eta_3$ given $\eta_2$ is zero, which corresponds to the conditional independence $\eta_1\perp\eta_3\mid\eta_2$. Three DAGs imply this relation without imposing additional marginal independencies, namely the chain $\eta_1\to\eta_2\to\eta_3$, the reverse chain $\eta_1\leftarrow\eta_2\leftarrow\eta_3$, and the fork $\eta_1\leftarrow\eta_2\to\eta_3$ (Figure~\ref{fig:mec_example}). Under the parameters shown, each reproduces $\Phi$ through \eqref{eq:latent_cov} and thus generates the same indicator distribution. In contrast, the DAG $\eta_1\to\eta_2\leftarrow\eta_3$ has both edges pointing into $\eta_2$, a collider. This DAG implies marginal independence $\eta_1\perp\eta_3$ ($\Phi_{13}=0$) and forms a distinct MEC. In this example, 
under the identification and faithfulness assumptions of Section~\ref{subsec:identifiability}, observational data identify the adjacencies in $\eta_1-\eta_2-\eta_3$ and rule out a collider at $\eta_2$, but they leave the direction of each edge undetermined.

Undirected graphical models \citep{lauritzen1996graphical}, including the partial correlation and Ising networks widely used in network psychometrics \citep{BorsboomCramer2013,MarsmanEtAl2018,EpskampWaldorpMottusBorsboom2018,BorsboomEtAl2021} and their latent extensions \citep{EpskampRhemtullaBorsboom2017}, encode a different family of conditional independence relations than directed graphs. By the pairwise Markov property, an absent edge implies conditional independence of the corresponding variables given all remaining variables, which is equivalent to zero partial correlation in the Gaussian setting \citep{Dempster1972}. For three variables, the undirected graph $\eta_1-\eta_2-\eta_3$ encodes $\eta_1\perp\eta_3\mid\eta_2$, corresponding to the MEC of the chain and the fork DAGs.
In contrast, the three-node DAG $\eta_1 \to \eta_2 \leftarrow \eta_3$ contains an unshielded collider, whose parents are not adjacent. It implies marginal independence ($\eta_1 \perp \eta_3$) but does not imply conditional independence given $\eta_2$, so no undirected graph encodes the same conditional independence relations.
Conversely, the four-cycle $\eta_1 - \eta_2 - \eta_3 - \eta_4 - \eta_1$ encodes two conditional independencies $\eta_1 \perp \eta_3 \mid (\eta_2, \eta_4)$ and $\eta_2 \perp \eta_4 \mid (\eta_1, \eta_3)$, a dependence structure that no DAG can represent faithfully \citep{Pearl1988,KollerFriedman2009}. 
In general, a DAG implies the same relations as an undirected graph when it has no unshielded collider, and an undirected graph implies the same relations as some DAG when every cycle of four or more vertices has a chord \citep{Wermuth1980,lauritzen1996graphical,AnderssonMadiganPerlman1997}. 
The two model classes therefore encode different dependence structures, and an estimated undirected network does not determine the directed structure that \eqref{eq:structural} targets \citep{RyanBringmannSchuurman2022}. 

\subsection{Prior Knowledge}\label{subsec:prior}

In many applications, domain expertise or experimental design provides partial prior knowledge about the directed structure. Typical examples include temporal ordering that forbids edges from later to earlier variables and substantive theory that rules out particular direct effects. We express such knowledge as zero restrictions on the path coefficient matrix $B$, which we can encode with a binary constraint matrix $\bZ\in\{0,1\}^{m\times m}$. Setting $Z_{ij}=0$ forces $\beta_{ij}=0$ and excludes the edge $\eta_i\to\eta_j$ a priori, whereas $Z_{ij}=1$ permits this edge and leaves $\beta_{ij}$ to be estimated from the data. We define the set of admissible DAGs compatible with $\bZ$ as
\[
  \calD_m(\bZ)=\{\calG(B)\in\calD_m:\beta_{ij}=0\text{ whenever }Z_{ij}=0\}.
\]

\begin{figure}[t]
  \centering
  \includegraphics[width=0.55\textwidth]{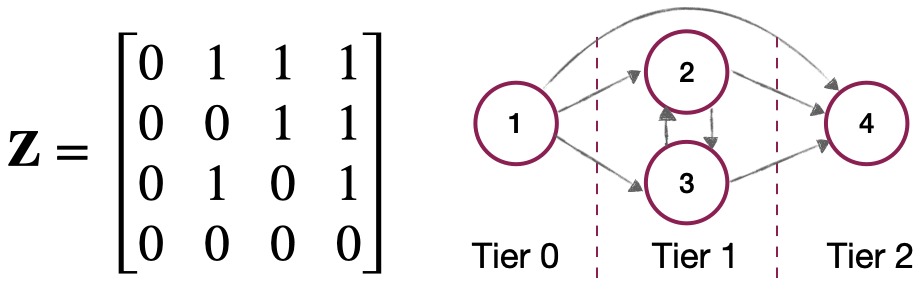}
  \caption{A constraint matrix $\bZ$ (left) and the candidate-edge diagram it encodes (right). The diagram imposes an ordering of tiers (Tier~0 $\to$ Tier~1 $\to$ Tier~2) and permits connections within Tier~1. Each link marks an entry $Z_{ij}=1$, a candidate edge whose presence and direction are learned from the data. The bidirectional pair between nodes~2 and~3 leaves that orientation open.}
  \label{fig:constraint_example}
\end{figure}

To illustrate, consider the constraint matrix in Figure~\ref{fig:constraint_example}, which enforces a hierarchical ordering of the latent variables (Tier~0 $\to$ Tier~1 $\to$ Tier~2) while permitting connections within Tier~1. 
In the candidate-edge diagram (Figure~\ref{fig:constraint_example}, right), each directed link marks an entry $Z_{ij}=1$, representing a potential edge.
When the constraint admits bidirectional candidates within a tier, for example $Z_{23}=1$ and $Z_{32}=1$, no directed cycle is imposed a priori. The direction is learned from the data, and the acyclicity constraint ensures that at most one of $\beta_{23}$ and $\beta_{32}$ is nonzero in the final estimate. 
Such background constraints can orient edges that would otherwise remain undirected within the Markov equivalence class.

\subsection{Learning Problem}\label{subsec:problem}

Let $\y_1,\ldots,\y_n\in\R^p$ be independent and identically distributed observations generated from the Gaussian linear SEM in \eqref{eq:measurement} and \eqref{eq:structural} with true parameters $\bftheta^*=(\Lambda^*,\Psi^*,B^*,\Omega^*)$, latent covariance matrix $\Phi^*=\Phi(B^*,\Omega^*)$, and indicator covariance matrix $\Sigma^*=\Sigma(\bftheta^*)$.

\begin{assumption}[Acyclic true structure]\label{ass:dag}
  Let $B^*$ be the true path coefficient matrix. Then there exists a true DAG $\calG^*=(\calV,\calE^*)$ such that $\beta_{ij}^*=0$ if and only if $(i,j)\notin\calE^*$.
\end{assumption}

We assume throughout that the prior knowledge is valid in that it excludes no true edges, so that $\calG^*\in\calD_m(\bZ)$. Because Markov equivalent DAGs induce identical observational distributions, the statistical target of inference is the MEC of $\calG^*$ restricted by $\bZ$, namely $\mathcal{M}(\mathcal{G}^*)\cap\mathcal{D}_m(\mathbf{Z})$, represented by $\CPDAG(\calG^*)$ together with $\bZ$. Under this setting, we address two central questions.

\begin{samepage}
\begin{itemize}
  \item[Q1.] Since the number of DAGs on $m$ vertices grows super-exponentially in $m$, how can this target be learned from the observed data in a computationally scalable manner?
  \item[Q2.] Is this target all that observational data can identify about $\calG^*$, and under what conditions does the procedure recover it consistently?
\end{itemize}
\end{samepage}
We address these questions in Sections~\ref{sec:alg} and \ref{sec:iden}, respectively.

\section{Proposed Method}\label{sec:alg}

To address question Q1, we propose DAGGLE (\textbf{D}irected \textbf{A}cyclic \textbf{G}raph learning in \textbf{G}aussian \textbf{L}inear structural \textbf{E}quation models), a two-stage estimation framework for general Gaussian linear SEMs \citep{Joreskog1973,BollenPearl2013}. DAGGLE follows the screening-and-pruning strategy of \citet{ZhangChenMoustaki2026}, with the exact Gaussian likelihood and the BIC under the measurement and structural restrictions of Section~\ref{sec:model}. 
DAGGLE seeks to minimize a Bayesian Information Criterion (BIC) over Markov equivalence classes. The criterion uses the marginal Gaussian likelihood of the observed indicators, with measurement and structural parameters estimated jointly.
Section~\ref{subsec:criterion} defines the likelihood fit function and the selection criterion, Sections~\ref{subsec:screening} and \ref{subsec:pruning} describe the two stages that minimize the criterion, and Section~\ref{subsec:alternatives} discusses alternative approaches, including the limits of two-step plug-in methods.

\subsection{Scoring Markov Equivalence Classes by BIC}\label{subsec:criterion}\label{subsec:daggle}

Let $\bS_n=n^{-1}\sum_{i=1}^n\y_i\y_i^\top$ be the sample covariance matrix of the centered indicators. The scaled negative Gaussian log-likelihood is
\begin{equation}\label{eq:nll}
  \ell_n(\bftheta)=\log|\Sigma(\bftheta)|+
  \tr\{\Sigma(\bftheta)^{-1}\bS_n\},
\end{equation}
which corresponds to the standard maximum likelihood discrepancy function in SEM up to an additive constant \citep{Joreskog1973,bollen1989structural}.

For an MEC with CPDAG $\calC$, we define
\begin{equation}\label{eq:bic}
  \mathrm{BIC}_n(\calC) = n\hat\ell_n(\calC) + d_{\calC}\log n.
\end{equation}
Here $\hat\ell_n(\calC)$ is the minimum of $\ell_n$ over the parameters whose path coefficients are zero outside the edges of a DAG in the MEC. Markov equivalent DAGs generate the same covariance matrices (Section~\ref{subsec:mec}), so this minimum does not depend on the DAG used for fitting (Section~\ref{subsec:consistency}). The penalty counts $d_{\calC}=d_{\mathrm{meas}}+|\calE(\calC)|$ free parameters, with $d_{\mathrm{meas}}$ from the measurement model $(\Lambda,\Psi)$ and one path coefficient for each of the $|\calE(\calC)|$ adjacencies in $\calC$. Deleting an adjacency thus lowers the penalty by $\log n$. Only MECs that contain a DAG in $\calD_m(\bZ)$ are candidates, and we call them admissible.

Minimizing $\mathrm{BIC}_n(\calC)$ directly over all admissible MECs faces two obstacles. First, the number of DAGs on $m$ vertices grows super-exponentially in $m$, which rules out an exhaustive search. Second, continuous optimization over the path coefficients avoids enumeration. However, the BIC cannot serve as its objective, because the penalty counts adjacencies and is not continuous in the coefficients. The penalty cannot be omitted either. Without it, any complete DAG yields a saturated structural model in which $\Phi$ is unrestricted and attains the smallest value of $\ell_n$.

DAGGLE therefore proceeds in two stages. In the first stage, we replace the discontinuous penalty with a continuous surrogate and minimize the resulting penalized objective over the parameters of all admissible DAGs at once. The second stage starts from the MEC obtained in the first and minimizes the BIC itself by a greedy backward search over neighboring MECs. The first stage thus screens for candidate MECs. It requires only that the selected MEC can reproduce the true covariance matrix $\Sigma^*$, even if it contains redundant edges. The second stage then serves as a pruning step that removes the redundant edges by BIC.

\subsection{Stage 1: Continuous Screening}\label{subsec:screening}\label{subsec:computation}

The first stage replaces the edge count in the BIC penalty by the truncated $\ell_1$ function $J_\tau$ \citep{shen2012likelihood}. Let $\Theta_{\bZ}$ denote the parameter space satisfying the restrictions of Section~\ref{subsec:model} and $\calG(B)\in\calD_m(\bZ)$. The screening problem is formulated as
\begin{equation}\label{eq:screen}
  \min_{\bftheta\in\Theta_{\bZ}}\Bigl\{\ell_n(\bftheta)+\lambda_n\sum_{i\neq j}J_{\tau_n}(\beta_{ij})\Bigr\},
  \qquad
  J_\tau(x)=\min\Bigl(\frac{|x|}{\tau},1\Bigr),
\end{equation}
with $\lambda_n=(\log n)/n$ and a truncation level $\tau_n>0$. Because $J_\tau(x) = |x|/\tau$ for $|x| < \tau$ and $J_\tau(x) = 1$ otherwise, the penalty $\sum_{i\neq j}J_\tau(\beta_{ij})$ provides a continuous approximation to the total number of directed edges in $\calG(B)$. Each coefficient with $|\beta_{ij}|\ge\tau_n$ then contributes $\log n$ to $n$ times the objective, matching the cost per free parameter under the BIC objective. The objective is thus a continuous surrogate of the BIC. Unlike the $\ell_1$ penalty, the truncated penalty remains constant beyond $\tau_n$, thereby avoiding shrinkage bias on larger effects. 

Two reformulations allow \eqref{eq:screen} to be solved by continuous optimization. First, following \citet{ZhengEtAl2018}, acyclicity is characterized by the smooth equality
\begin{equation}\label{eq:acyclicity}
  h(B)=\tr\{\exp(B\circ B)\}-m=0,
\end{equation}
where $\circ$ denotes the Hadamard product and $\exp(\cdot)$ the matrix exponential. For any matrix $B$ with a zero diagonal, $h(B)=0$ if and only if $\calG(B)$ is acyclic. The combinatorial search over DAGs thus becomes a smooth equality-constrained problem. Alternatively, one can employ the log-determinant characterization developed by \citet{bello2022dagma}. The prior restrictions enter as structural zeros through $B=\bZ\circ W$ with an unrestricted $W\in\R^{m\times m}$. Second, the penalty admits the decomposition $J_\tau(x)=|x|/\tau-\max(|x|/\tau-1,0)$ as a difference of two convex functions. We apply a difference-of-convex algorithm \citep{shen2012likelihood} that linearizes the second convex component at the current iterate. Setting $A_t=\{(i,j):|\beta^{(t)}_{ij}|\le\tau_n\}$, the update at iteration $t+1$ solves the weighted $\ell_1$ subproblem
\begin{equation}\label{eq:optim_continuous}
  \min_{\bftheta}\ \ell_n(\bftheta)+\frac{\lambda_n}{\tau_n}\sum_{(i,j)\in A_t}|\beta_{ij}|
  \quad\text{subject to}\quad h(B)=0\ \text{ and }\ B=\bZ\circ W
\end{equation}
over the parameter values that satisfy the restrictions of Section~\ref{subsec:model}. Only the coefficients below the truncation threshold are penalized. We solve each subproblem using the augmented Lagrangian method \citep[Chapter~17]{nocedal2006numerical}, and the iterations stop when the active set $A_t$ stabilizes.

Let $\tilde\bftheta_n$ denote a computed solution of \eqref{eq:screen} with coefficient matrix $\tilde B_n$. Defining $\calG_\tau(B)$ as the directed graph with edge set $\{(i,j):|\beta_{ij}|\ge\tau\}$, the screened DAG is $\calG_{\tau_n}(\tilde B_n)$. 
Because the problem is non-convex, it is evaluated across multiple random initializations and over a decreasing sequence of truncation levels. 
Each candidate graph is then refitted without penalty, and the first stage returns the one whose MEC minimizes the BIC. 
We denote the selected DAG by $\hat\calG_n^{(0)}$, its unpenalized estimate by $\hat\bftheta_n^{(0)}$, and its MEC by $\hat\calC_n^{(0)}=\CPDAG(\hat\calG_n^{(0)})$. Because numerical solutions cannot guarantee global optimality, the initial screened MEC may retain redundant adjacencies, as shown in Section~\ref{sec:simulation}. The second stage prunes these remaining false discoveries.

\subsection{Stage 2: Backward Pruning over Markov Equivalence Classes}\label{subsec:pruning}

The second stage starts from $\hat\calC_n^{(0)}$ and performs a greedy backward search over admissible MECs. For a CPDAG $\calC$, let $[\calC]_{\bZ}=\{\calG\in\calD_m(\bZ):\CPDAG(\calG)=\calC\}$ denote the set of its admissible DAG representations. We define the backward deletion neighborhood of $\calC$ as
\begin{equation}\label{eq:cpdag_neighbors}
  \calN^-_{\bZ}(\calC)
  =\{\CPDAG(\calG\setminus e):\calG\in[\calC]_{\bZ},\ e\text{ an edge of }\calG\}.
\end{equation}
Deletions are taken from every admissible member of the MEC, because the MEC reached by a deletion depends on how the member orients the remaining edges. The neighborhood is generated by the delete operator of GES \citep{chickering2002equivalence,chickering2002optimal,meek1995causal}. The search iteratively steps to the neighbor that most reduces the BIC until no neighbor improves the BIC (Algorithm~\ref{alg:two_step}). Each neighbor $\calC'$ is nested in $\calC$ with one fewer parameter, so it improves on $\calC$ if and only if $n\{\hat\ell_n(\calC')-\hat\ell_n(\calC)\}<\log n$. Each move is thus a likelihood ratio comparison of nested models with critical value $\log n$.

\begin{algorithm}
  \caption{Two-stage estimation in DAGGLE}\label{alg:two_step}
  \begin{algorithmic}[1]
    \Require Data $\y_1,\ldots,\y_n$, matrices $Q$ and $\bZ$, starting values, and truncation levels.
    \For{each starting value and truncation level $\tau_n$}\Comment{Stage 1}
      \State Solve the continuous screening problem \eqref{eq:screen}, apply threshold $\tau_n$, and evaluate the BIC of the resulting MEC.
    \EndFor
    \State Set $\calC\leftarrow\hat\calC^{(0)}_n$, the screened MEC that minimizes the BIC
    \While{there exists $\calC'\in\calN^-_{\bZ}(\calC)$ satisfying $\mathrm{BIC}_n(\calC')<\mathrm{BIC}_n(\calC)$}\Comment{Stage 2}
      \State $\calC\leftarrow\argmin_{\calC'\in\calN^-_{\bZ}(\calC)}\mathrm{BIC}_n(\calC')$
    \EndWhile
    \State \Return $\hat\calC_n=\calC$
  \end{algorithmic}
\end{algorithm}

When the current Markov equivalence class is dense, the deletion neighborhood can grow exponentially with the number of latent variables $m$ \citep{chickering2002optimal}. Initiating backward elimination directly from the complete graph, as practiced in some latent-variable procedures \citep{DongEtAl2026LGES}, is therefore computationally prohibitive. Initializing the search at the screened equivalence class overcomes this difficulty by ensuring that the deletion neighborhoods remain small whenever the preliminary screen is sparse.

\subsection{Alternative Methods}\label{subsec:alternatives}\label{subsec:factor_scores}

An alternative two-step method first fits the measurement model and then applies PC \citep{SpirtesGlymourScheines2000} or GES \citep{chickering2002optimal} to factor scores or an estimate of $\Phi$. However, factor-score regressions can be biased \citep{SkrondalLaake2001,Croon2002,DevliegerMayerRosseel2016}; for graph learning, a further concern is whether proxies preserve the latent conditional independences.

\begin{samepage}
\begin{example}[Conditional independence under measurement error]\label{ex:proxy_ci_distortion}
  Consider a standardized Gaussian chain $\eta_1\to\eta_2\to\eta_3$ with $\Cov(\eta_1,\eta_2)=a$, $\Cov(\eta_2,\eta_3)=b$, and $ab\neq0$.
  Let $\tilde\eta_2=\eta_2+\epsilon$, where $\epsilon\sim\calN(0,\kappa^2)$ is independent of the latent variables and $\kappa^2>0$.
  Although $\eta_1\perp\eta_3\mid\eta_2$,
  \[
    \Cov(\eta_1,\eta_3\mid\tilde\eta_2)
    =\frac{ab\kappa^2}{1+\kappa^2}\neq0.
  \]
\end{example}
\end{samepage}

Regression factor scores are conditional means rather than proxies with independent additive error, but they can also alter conditional independences. Section~S1.1 gives a population counterexample and the general covariance identity underlying Example~\ref{ex:proxy_ci_distortion}.

A structure learner applied to a consistent estimate of $\Phi$ can target the latent conditional independence structure. The uncertainty in this estimate still depends on measurement quality: the plug-in error bound in Section~S1.1 grows as loadings weaken, even with known measurement parameters. Treating the estimate as an ordinary sample covariance does not generally account for its sampling distribution. DAGGLE incorporates the measurement model through \eqref{eq:nll}, but remains affected by weak loadings.

A second alternative selects edges by thresholding an $\ell_1$-penalized fit, as in NOTEARS \citep{ZhengEtAl2018} and DAGMA \citep{bello2022dagma}. The selected graph depends on both the penalty level and the threshold. DAGGLE uses its penalized fit for screening and selects the final MEC by backward pruning with the BIC, refitting each candidate without penalization. Section~\ref{sec:simulation} compares DAGGLE with the plug-in methods and its unpruned screen.

\section{Theoretical Properties}\label{sec:iden}

This section addresses question Q2 by developing the theoretical foundations of the proposed framework. Section~\ref{subsec:identifiability} establishes that the indicators identify the MEC of the true DAG $\calG^*$ subject to valid prior knowledge. Section~\ref{subsec:consistency} then discusses the conditions under which DAGGLE recovers this MEC consistently. Proofs are deferred to Section~S1.

\subsection{Identifiability}\label{subsec:identifiability}

Identification proceeds in two steps. The indicators first identify the measurement parameters and the latent covariance matrix. The latent covariance matrix then identifies the graph up to Markov equivalence, as in DAG learning with fully observed variables \citep{verma1990equivalence,pearl2009causality}. We formalize the measurement identification requirement below.

\begin{assumption}[Identification of the measurement model]\label{ass:mea_iden}
  The true matrices $(\Lambda^*,\Phi^*,\Psi^*)$ satisfy the measurement restrictions and normalizations of Section~\ref{subsec:model}, with $\Phi^*$ and $\Psi^*$ positive definite. If a triple $(\Lambda^\dagger,\Phi^\dagger,\Psi^\dagger)$ satisfying the same restrictions has
  \[
    \Lambda^\dagger\Phi^\dagger(\Lambda^\dagger)^\top+\Psi^\dagger
    =\Lambda^*\Phi^*(\Lambda^*)^\top+\Psi^*,
  \]
  then $(\Lambda^\dagger,\Phi^\dagger,\Psi^\dagger) = (\Lambda^*,\Phi^*,\Psi^*)$.
\end{assumption}

\begin{remark}[Measurement model constraints]\label{rem:measurement}
  The decomposition $\Lambda\Phi\Lambda^\top+\Psi$ is unchanged when $(\Lambda,\Phi)$ is replaced by $(\Lambda M,M^{-1}\Phi M^{-\top})$ for an invertible $M$, so without restrictions the latent variables are identified only up to rotation, scale, and sign. As in confirmatory factor analysis, the pattern $Q$ and the normalizations remove this indeterminacy. Apart from its unit diagonal, $\Phi$ is left unrestricted, because it is induced by the unknown DAG. A simple sufficient condition is the \emph{three-indicator rule}, with diagonal $\Psi$ and at least three indicators loading on each latent variable alone \citep[Chapter~7]{bollen1989structural}. Two indicators suffice when each latent variable is correlated with another. For more general loading patterns with diagonal $\Psi$, the row criterion of \citet[Theorem~5.1]{anderson1956statistical} identifies $\Psi$, and suitable loading restrictions resolve the rotation.
\end{remark}
 
Identifying the graph from the latent covariance matrix requires faithfulness. A distribution of $\bfeta$ is faithful to a DAG $\calG$ if every conditional independence it satisfies is implied by $\calG$ (Definition~\ref{def:mec}). Faithfulness is a generic property, since the set of unfaithful path coefficients has Lebesgue measure zero in the parameter space (\citealp[Theorem~8]{meek1995strong}; \citealp{SpirtesGlymourScheines2000}).

\begin{assumption}[Faithfulness]\label{ass:faithful}
  The latent distribution $\calN(\bzero,\Phi^*)$ is faithful to $\calG^*$.
\end{assumption}

\begin{theorem}[Identifiability of the latent DAG]\label{thm:latent-ident}
  Suppose Assumptions~\ref{ass:dag}--\ref{ass:faithful} hold.
  \begin{enumerate}
    \item[\textup{(i)}] If $\bftheta^\dagger=(\Lambda^\dagger,\Psi^\dagger,B^\dagger,\Omega^\dagger)$ satisfies the same measurement constraints and normalizations, has a diagonal positive definite $\Omega^\dagger$ and a DAG $\calG(B^\dagger)$ to which its latent distribution is faithful, and satisfies $\Sigma(\bftheta^\dagger)=\Sigma(\bftheta^*)$, then $\calG(B^\dagger)\in\calM(\calG^*)$.
    \item[\textup{(ii)}] Conversely, for every DAG $\calG^\dagger\in\calM(\calG^*)$, there exist $B^\dagger$ and a diagonal positive definite $\Omega^\dagger$ with $\calG(B^\dagger)=\calG^\dagger$ such that $\bftheta^\dagger=(\Lambda^*,\Psi^*,B^\dagger,\Omega^\dagger)$ satisfies $\Sigma(\bftheta^\dagger)=\Sigma(\bftheta^*)$.
  \end{enumerate}
\end{theorem}

Theorem~\ref{thm:latent-ident} shows that $\calM(\calG^*)$ is the sharpest structural object identifiable from observational data. It determines the adjacencies and every edge direction shared by all members of the MEC, but generally not $\calG^*$ itself. Prior knowledge of the form in Section~\ref{subsec:prior} restricts the identified MEC further.

\begin{corollary}[Identifiability under prior knowledge]\label{cor:prior_knowledge}
  Suppose Assumptions~\ref{ass:dag}--\ref{ass:faithful} hold and $\calG^*\in\calD_m(\bZ)$. Then $\calG^*$ is identified up to the restricted MEC $\calM(\calG^*)\cap\calD_m(\bZ)$, that is, the DAGs that are observationally equivalent to $\calG^*$ and compatible with $\bZ$. In particular, $\calG^*$ is identified when $\bZ$ excludes every other member of $\calM(\calG^*)$.
\end{corollary}

Prior knowledge thus acts by orienting reversible edges of the MEC, and DAGGLE uses it by optimizing over $\calD_m(\bZ)$.

\subsection{Consistency of DAGGLE}\label{subsec:consistency}\label{subsec:theory}

Under $\diag(\Phi)=\bone_m$, a DAG generates the positive definite correlation matrices satisfying the conditional independence relations it implies (proof of Theorem~\ref{thm:latent-ident}(ii)). Markov equivalent DAGs therefore have the same indicator covariance model and minimum of $\ell_n$. Since the normalization determines $\Omega$ from $B$, each adjacency contributes one free parameter. Hence the BIC in \eqref{eq:bic} is invariant within an MEC.
 
For fixed $m$ and $p$, DAGGLE recovers the true Markov equivalence class with probability tending to one as $n\to\infty$. This result holds under Assumptions~\ref{ass:dag}--\ref{ass:faithful}, Gaussian likelihood regularity, and sufficient numerical accuracy in both stages. Furthermore, we assume that the prior restrictions are valid and define ordered tiers (possibly with forbidden adjacencies), where edges within the same tier may take either direction.

Detailed assumptions and proofs are provided in Sections~S1.3--S1.4 of the Supplementary Material. The proof adapts the consistency argument of \citet[Theorem~1]{ZhangChenMoustaki2026} to the exact Gaussian marginal likelihood with BIC and these prior restrictions. Screening retains a correctly specified model with probability tending to one, after which BIC-guided pruning removes redundant edges to recover the true class.

\section{Simulation Study}\label{sec:simulation}

This simulation study evaluates graph recovery accuracy, evaluates the specific empirical contribution of backward pruning, and compares DAGGLE with established plug-in alternatives.

\subsection{Design and Evaluation}\label{subsec:sim_design}

The true DAGs on $m$ latent variables are generated from three families. (1) \textbf{Random} graphs have $m$ edges on uniformly sampled node pairs. Each edge points from the earlier to the later node in a random permutation of the nodes, which guarantees acyclicity. (2) \textbf{Star} graphs consist of a randomly chosen hub that receives edges from $\lceil(m-1)/2\rceil$ randomly selected nodes, so every edge of the true CPDAG is directed. (3) \textbf{Tree} graphs form a balanced binary tree at $m=20$ and a balanced ternary tree at $m=40$, with edges directed outward from the root, so the true CPDAG is undirected. Nonzero path coefficients are sampled from $U(0.5,2.0)$ with random signs. Each latent variable has three continuous indicators. On the original scale, the first loading in each block equals one, the remaining loadings are drawn independently from $U(1.0,2.0)$, and the measurement-error variances equal one. The loadings and path coefficients are then rescaled so that every latent variable has unit variance as in Section~\ref{subsec:model}, which leaves the graph and its CPDAG unchanged. Crossing the three families with latent dimension $m\in\{20,40\}$ and sample size $n\in\{2000,4000\}$ gives $12$ settings with $p=3m$ indicators. Each condition comprises $100$ replications and uses the same data for all methods.

We evaluate five methods. (1) \textbf{DAGGLE} uses the Gaussian likelihood in both stages, with penalty weight $\lambda_n=(\log n)/n$, five random starts, and truncation levels $\tau_n=0.10$, $0.05$, and $0.02$. Each candidate screen is refitted without penalty, and the MEC with the smallest BIC initializes pruning. (2) \textbf{DAGGLE$^{(0)}$} returns this unpruned initial CPDAG and isolates the contribution of the Stage 2 pruning. The remaining competitors are the two-step plug-in procedures. Each initially estimates the confirmatory factor model by maximum likelihood using the true configuration matrix $Q$ and an unrestricted latent correlation matrix. (3) \textbf{FS-PC} applies the PC algorithm to regression factor scores, with Fisher-$z$ tests at level $0.01$ and conditioning sets of at most four variables. (4) \textbf{FS-GES} applies GES with the Gaussian BIC to the same scores. (5) \textbf{LC-GES} applies GES to the estimated latent correlation matrix, treated as the sample covariance matrix of $n$ Gaussian observations.

Graph recovery is evaluated by comparing the estimated CPDAG with the true CPDAG through five criteria. The structural Hamming distance (SHD) counts the node pairs whose adjacency or orientation differs. Skeleton F1 is the harmonic mean of precision and recall for adjacencies. Directed-edge F1 is defined in the same way for directed edges. Exact recovery (MEC) is the proportion of replications in which the two CPDAGs coincide, and Extra is the mean number of false-positive adjacencies. Formal definitions of these criteria appear in Section~S3. Directed-edge F1 is omitted for trees, whose true CPDAGs have no directed edges. Each method has an $8$-hour budget per replication, and summaries for incomplete settings use the completed replications.
 
\begin{table}[t]
  \centering
  \scriptsize
  \singlespacing
  \caption{Recovery of the latent MEC under the linear factor model at $m=20$, based on $100$ replications per setting. The columns give mean CPDAG structural Hamming distance (SHD), skeleton F1 (Skel.), directed-edge F1 (Arrow), exact-recovery rate (MEC), and mean number of false-positive adjacencies (Extra). Monte Carlo standard errors are in parentheses. Bold marks the best value in each column within a panel, including ties at the displayed precision. Arrow F1 is not reported for trees, whose true CPDAGs are undirected.}
  \label{tab:gaussian_recovery_m20}
  \setlength{\tabcolsep}{2pt}
  \resizebox{\linewidth}{!}{\begin{tabular}{l rrrrr rrrrr}
    \toprule
             & \multicolumn{5}{c}{$n{=}2000$}   & \multicolumn{5}{c}{$n{=}4000$} \\
    \cmidrule(lr){2-6}\cmidrule(lr){7-11}
    Method & \multicolumn{1}{c}{SHD} & \multicolumn{1}{c}{Skel.} & \multicolumn{1}{c}{Arrow} & \multicolumn{1}{c}{MEC} & \multicolumn{1}{c}{Extra} & \multicolumn{1}{c}{SHD} & \multicolumn{1}{c}{Skel.} & \multicolumn{1}{c}{Arrow} & \multicolumn{1}{c}{MEC} & \multicolumn{1}{c}{Extra} \\
    \midrule
    \multicolumn{11}{l}{\emph{Panel A. Random}} \\
    DAGGLE & \textbf{2.3}\,{\scriptsize(0.4)} & \textbf{0.97}\,{\scriptsize(0.00)} & \textbf{0.93}\,{\scriptsize(0.01)} & \textbf{0.46}\,{\scriptsize(0.05)} & \textbf{1.3}\,{\scriptsize(0.2)} & \textbf{0.8}\,{\scriptsize(0.2)} & \textbf{0.99}\,{\scriptsize(0.00)} & \textbf{0.98}\,{\scriptsize(0.01)} & \textbf{0.69}\,{\scriptsize(0.05)} & \textbf{0.4}\,{\scriptsize(0.1)} \\
    DAGGLE$^{(0)}$ & 8.8\,{\scriptsize(0.8)} & 0.88\,{\scriptsize(0.01)} & 0.79\,{\scriptsize(0.02)} & 0.12\,{\scriptsize(0.03)} & 6.1\,{\scriptsize(0.7)} & 5.2\,{\scriptsize(0.5)} & 0.93\,{\scriptsize(0.01)} & 0.87\,{\scriptsize(0.01)} & 0.24\,{\scriptsize(0.04)} & 3.2\,{\scriptsize(0.3)} \\
    FS-PC & 16.8\,{\scriptsize(0.6)} & 0.79\,{\scriptsize(0.01)} & 0.51\,{\scriptsize(0.02)} & 0.00\,{\scriptsize(0.00)} & 8.8\,{\scriptsize(0.3)} & 18.8\,{\scriptsize(0.5)} & 0.75\,{\scriptsize(0.01)} & 0.49\,{\scriptsize(0.01)} & 0.00\,{\scriptsize(0.00)} & 11.1\,{\scriptsize(0.3)} \\
    FS-GES & 47.0\,{\scriptsize(0.9)} & 0.50\,{\scriptsize(0.01)} & 0.34\,{\scriptsize(0.01)} & 0.00\,{\scriptsize(0.00)} & 40.0\,{\scriptsize(0.8)} & 45.2\,{\scriptsize(0.7)} & 0.50\,{\scriptsize(0.00)} & 0.37\,{\scriptsize(0.01)} & 0.00\,{\scriptsize(0.00)} & 39.5\,{\scriptsize(0.6)} \\
    LC-GES & 18.8\,{\scriptsize(0.7)} & 0.76\,{\scriptsize(0.01)} & 0.56\,{\scriptsize(0.01)} & 0.00\,{\scriptsize(0.00)} & 12.9\,{\scriptsize(0.5)} & 14.2\,{\scriptsize(0.5)} & 0.80\,{\scriptsize(0.01)} & 0.66\,{\scriptsize(0.01)} & 0.00\,{\scriptsize(0.00)} & 10.2\,{\scriptsize(0.4)} \\
    \midrule
    \multicolumn{11}{l}{\emph{Panel B. Star}} \\
    DAGGLE & \textbf{1.0}\,{\scriptsize(0.1)} & \textbf{0.95}\,{\scriptsize(0.00)} & \textbf{0.99}\,{\scriptsize(0.00)} & \textbf{0.31}\,{\scriptsize(0.05)} & \textbf{1.0}\,{\scriptsize(0.1)} & \textbf{0.6}\,{\scriptsize(0.1)} & \textbf{0.97}\,{\scriptsize(0.00)} & \textbf{1.00}\,{\scriptsize(0.00)} & \textbf{0.55}\,{\scriptsize(0.05)} & \textbf{0.6}\,{\scriptsize(0.1)} \\
    DAGGLE$^{(0)}$ & 1.2\,{\scriptsize(0.2)} & \textbf{0.95}\,{\scriptsize(0.01)} & \textbf{0.99}\,{\scriptsize(0.01)} & 0.30\,{\scriptsize(0.05)} & 1.2\,{\scriptsize(0.2)} & 0.9\,{\scriptsize(0.2)} & 0.96\,{\scriptsize(0.01)} & 0.99\,{\scriptsize(0.00)} & \textbf{0.55}\,{\scriptsize(0.05)} & 0.9\,{\scriptsize(0.1)} \\
    FS-PC & 10.1\,{\scriptsize(0.3)} & 0.70\,{\scriptsize(0.01)} & 0.65\,{\scriptsize(0.01)} & 0.00\,{\scriptsize(0.00)} & 8.7\,{\scriptsize(0.3)} & 10.5\,{\scriptsize(0.3)} & 0.69\,{\scriptsize(0.01)} & 0.65\,{\scriptsize(0.01)} & 0.00\,{\scriptsize(0.00)} & 9.2\,{\scriptsize(0.3)} \\
    FS-GES & 14.3\,{\scriptsize(0.3)} & 0.59\,{\scriptsize(0.01)} & 0.65\,{\scriptsize(0.01)} & 0.00\,{\scriptsize(0.00)} & 14.3\,{\scriptsize(0.3)} & 12.2\,{\scriptsize(0.3)} & 0.63\,{\scriptsize(0.01)} & 0.70\,{\scriptsize(0.01)} & 0.00\,{\scriptsize(0.00)} & 12.2\,{\scriptsize(0.3)} \\
    LC-GES & 5.9\,{\scriptsize(0.2)} & 0.78\,{\scriptsize(0.01)} & 0.85\,{\scriptsize(0.01)} & 0.01\,{\scriptsize(0.01)} & 5.9\,{\scriptsize(0.2)} & 4.7\,{\scriptsize(0.2)} & 0.82\,{\scriptsize(0.01)} & 0.89\,{\scriptsize(0.01)} & 0.01\,{\scriptsize(0.01)} & 4.7\,{\scriptsize(0.2)} \\
    \midrule
    \multicolumn{11}{l}{\emph{Panel C. Tree}} \\
    DAGGLE & \textbf{2.0}\,{\scriptsize(0.4)} & \textbf{0.98}\,{\scriptsize(0.00)} & \multicolumn{1}{c}{--} & \textbf{0.63}\,{\scriptsize(0.05)} & \textbf{0.8}\,{\scriptsize(0.1)} & \textbf{2.2}\,{\scriptsize(0.5)} & \textbf{0.98}\,{\scriptsize(0.00)} & \multicolumn{1}{c}{--} & \textbf{0.71}\,{\scriptsize(0.05)} & \textbf{0.7}\,{\scriptsize(0.2)} \\
    DAGGLE$^{(0)}$ & 8.6\,{\scriptsize(0.5)} & 0.86\,{\scriptsize(0.01)} & \multicolumn{1}{c}{--} & 0.02\,{\scriptsize(0.01)} & 6.2\,{\scriptsize(0.3)} & 6.6\,{\scriptsize(0.5)} & 0.89\,{\scriptsize(0.01)} & \multicolumn{1}{c}{--} & 0.01\,{\scriptsize(0.01)} & 4.8\,{\scriptsize(0.3)} \\
    FS-PC & 31.1\,{\scriptsize(0.4)} & 0.70\,{\scriptsize(0.00)} & \multicolumn{1}{c}{--} & 0.00\,{\scriptsize(0.00)} & 16.5\,{\scriptsize(0.2)} & 35.7\,{\scriptsize(0.4)} & 0.65\,{\scriptsize(0.00)} & \multicolumn{1}{c}{--} & 0.00\,{\scriptsize(0.00)} & 20.4\,{\scriptsize(0.2)} \\
    FS-GES & 54.9\,{\scriptsize(0.7)} & 0.50\,{\scriptsize(0.00)} & \multicolumn{1}{c}{--} & 0.00\,{\scriptsize(0.00)} & 38.6\,{\scriptsize(0.7)} & 54.4\,{\scriptsize(0.6)} & 0.50\,{\scriptsize(0.00)} & \multicolumn{1}{c}{--} & 0.00\,{\scriptsize(0.00)} & 38.6\,{\scriptsize(0.6)} \\
    LC-GES & 17.5\,{\scriptsize(0.6)} & 0.83\,{\scriptsize(0.01)} & \multicolumn{1}{c}{--} & 0.00\,{\scriptsize(0.00)} & 7.7\,{\scriptsize(0.3)} & 14.6\,{\scriptsize(0.6)} & 0.87\,{\scriptsize(0.01)} & \multicolumn{1}{c}{--} & 0.01\,{\scriptsize(0.01)} & 6.1\,{\scriptsize(0.3)} \\
    \bottomrule
  \end{tabular}}
\end{table}

\subsection{Recovery Results}\label{subsec:sim_results}

Tables~\ref{tab:gaussian_recovery_m20} and~\ref{tab:gaussian_recovery_m40} summarize graph recovery at $m=20$ and $m=40$.
DAGGLE attains the lowest mean SHD and the highest exact-recovery rate in all $12$ settings, and its advantage over the plug-in methods is substantial.
The plug-in methods rarely recover the exact MEC, whereas DAGGLE does so in a sizable proportion of replications in most settings.
Its remaining errors have a simple form.
On star graphs, nearly every remaining error is an extra adjacency; on random graphs, extra adjacencies account for at least half of the SHD, and the directed-edge F1 exceeds $0.9$ wherever it is defined, so on random and star graphs DAGGLE occasionally retains a spurious edge but orients nearly all the edges it keeps correctly.
Errors grow with $m$ and are largest on the tree graphs at $m=40$, where the skeleton F1 remains above $0.9$ and the errors concern orientation.
On random and star graphs a larger sample reduces the errors at both dimensions, consistent with the asymptotic result of Section~\ref{subsec:consistency}.

\begin{table}[t]
  \centering
  \scriptsize
  \singlespacing
  \caption{Recovery of the latent MEC under the linear factor model at $m=40$, based on $100$ replications per cell, with the columns of Table~\ref{tab:gaussian_recovery_m20}. FS-GES did not finish every replication on tree graphs within the $8$-hour budget. The numbers of replications finished at $n{=}2000$ and $n{=}4000$ follow its name, and its entries average these replications.}
  \label{tab:gaussian_recovery_m40}
  \setlength{\tabcolsep}{2pt}
  \resizebox{\linewidth}{!}{\begin{tabular}{l rrrrr rrrrr}
    \toprule
             & \multicolumn{5}{c}{$n{=}2000$}   & \multicolumn{5}{c}{$n{=}4000$} \\
    \cmidrule(lr){2-6}\cmidrule(lr){7-11}
    Method & \multicolumn{1}{c}{SHD} & \multicolumn{1}{c}{Skel.} & \multicolumn{1}{c}{Arrow} & \multicolumn{1}{c}{MEC} & \multicolumn{1}{c}{Extra} & \multicolumn{1}{c}{SHD} & \multicolumn{1}{c}{Skel.} & \multicolumn{1}{c}{Arrow} & \multicolumn{1}{c}{MEC} & \multicolumn{1}{c}{Extra} \\
    \midrule
    \multicolumn{11}{l}{\emph{Panel A. Random}} \\
    DAGGLE & \textbf{5.2}\,{\scriptsize(0.8)} & \textbf{0.96}\,{\scriptsize(0.01)} & \textbf{0.92}\,{\scriptsize(0.01)} & \textbf{0.47}\,{\scriptsize(0.05)} & \textbf{3.4}\,{\scriptsize(0.6)} & \textbf{3.3}\,{\scriptsize(0.7)} & \textbf{0.98}\,{\scriptsize(0.01)} & \textbf{0.95}\,{\scriptsize(0.01)} & \textbf{0.62}\,{\scriptsize(0.05)} & \textbf{2.2}\,{\scriptsize(0.5)} \\
    DAGGLE$^{(0)}$ & 29.6\,{\scriptsize(2.4)} & 0.80\,{\scriptsize(0.01)} & 0.70\,{\scriptsize(0.02)} & 0.01\,{\scriptsize(0.01)} & 22.9\,{\scriptsize(2.2)} & 20.1\,{\scriptsize(1.4)} & 0.86\,{\scriptsize(0.01)} & 0.75\,{\scriptsize(0.01)} & 0.01\,{\scriptsize(0.01)} & 14.1\,{\scriptsize(1.2)} \\
    FS-PC & 37.6\,{\scriptsize(0.7)} & 0.76\,{\scriptsize(0.00)} & 0.49\,{\scriptsize(0.01)} & 0.00\,{\scriptsize(0.00)} & 21.4\,{\scriptsize(0.4)} & 41.1\,{\scriptsize(0.8)} & 0.74\,{\scriptsize(0.00)} & 0.47\,{\scriptsize(0.01)} & 0.00\,{\scriptsize(0.00)} & 25.1\,{\scriptsize(0.4)} \\
    FS-GES & 151.9\,{\scriptsize(2.0)} & 0.36\,{\scriptsize(0.00)} & 0.27\,{\scriptsize(0.00)} & 0.00\,{\scriptsize(0.00)} & 140.5\,{\scriptsize(1.9)} & 146.2\,{\scriptsize(1.7)} & 0.37\,{\scriptsize(0.00)} & 0.28\,{\scriptsize(0.00)} & 0.00\,{\scriptsize(0.00)} & 134.4\,{\scriptsize(1.7)} \\
    LC-GES & 54.0\,{\scriptsize(0.9)} & 0.64\,{\scriptsize(0.00)} & 0.52\,{\scriptsize(0.01)} & 0.00\,{\scriptsize(0.00)} & 44.0\,{\scriptsize(0.8)} & 47.8\,{\scriptsize(1.0)} & 0.68\,{\scriptsize(0.00)} & 0.54\,{\scriptsize(0.01)} & 0.00\,{\scriptsize(0.00)} & 37.6\,{\scriptsize(0.8)} \\
    \midrule
    \multicolumn{11}{l}{\emph{Panel B. Star}} \\
    DAGGLE & \textbf{3.3}\,{\scriptsize(0.2)} & \textbf{0.92}\,{\scriptsize(0.00)} & \textbf{0.99}\,{\scriptsize(0.00)} & \textbf{0.08}\,{\scriptsize(0.03)} & \textbf{3.2}\,{\scriptsize(0.2)} & \textbf{1.8}\,{\scriptsize(0.2)} & \textbf{0.96}\,{\scriptsize(0.00)} & \textbf{1.00}\,{\scriptsize(0.00)} & \textbf{0.24}\,{\scriptsize(0.04)} & \textbf{1.8}\,{\scriptsize(0.2)} \\
    DAGGLE$^{(0)}$ & 10.0\,{\scriptsize(0.9)} & 0.82\,{\scriptsize(0.01)} & 0.89\,{\scriptsize(0.01)} & 0.06\,{\scriptsize(0.02)} & 9.7\,{\scriptsize(0.9)} & 4.8\,{\scriptsize(0.7)} & 0.91\,{\scriptsize(0.01)} & 0.95\,{\scriptsize(0.01)} & 0.21\,{\scriptsize(0.04)} & 4.7\,{\scriptsize(0.7)} \\
    FS-PC & 36.8\,{\scriptsize(0.6)} & 0.55\,{\scriptsize(0.00)} & 0.44\,{\scriptsize(0.01)} & 0.00\,{\scriptsize(0.00)} & 31.7\,{\scriptsize(0.5)} & 37.8\,{\scriptsize(0.7)} & 0.55\,{\scriptsize(0.00)} & 0.45\,{\scriptsize(0.01)} & 0.00\,{\scriptsize(0.00)} & 33.3\,{\scriptsize(0.6)} \\
    FS-GES & 51.4\,{\scriptsize(0.6)} & 0.44\,{\scriptsize(0.00)} & 0.47\,{\scriptsize(0.00)} & 0.00\,{\scriptsize(0.00)} & 51.4\,{\scriptsize(0.6)} & 43.4\,{\scriptsize(0.6)} & 0.48\,{\scriptsize(0.00)} & 0.53\,{\scriptsize(0.00)} & 0.00\,{\scriptsize(0.00)} & 43.4\,{\scriptsize(0.6)} \\
    LC-GES & 24.2\,{\scriptsize(0.5)} & 0.63\,{\scriptsize(0.00)} & 0.70\,{\scriptsize(0.01)} & 0.00\,{\scriptsize(0.00)} & 24.2\,{\scriptsize(0.5)} & 18.7\,{\scriptsize(0.5)} & 0.69\,{\scriptsize(0.01)} & 0.77\,{\scriptsize(0.01)} & 0.00\,{\scriptsize(0.00)} & 18.7\,{\scriptsize(0.5)} \\
    \midrule
    \multicolumn{11}{l}{\emph{Panel C. Tree}} \\
    DAGGLE & \textbf{18.1}\,{\scriptsize(1.8)} & \textbf{0.94}\,{\scriptsize(0.01)} & \multicolumn{1}{c}{--} & \textbf{0.25}\,{\scriptsize(0.04)} & \textbf{5.3}\,{\scriptsize(0.7)} & \textbf{19.0}\,{\scriptsize(1.9)} & \textbf{0.94}\,{\scriptsize(0.01)} & \multicolumn{1}{c}{--} & \textbf{0.32}\,{\scriptsize(0.05)} & \textbf{5.6}\,{\scriptsize(0.7)} \\
    DAGGLE$^{(0)}$ & 43.8\,{\scriptsize(2.3)} & 0.73\,{\scriptsize(0.01)} & \multicolumn{1}{c}{--} & 0.00\,{\scriptsize(0.00)} & 29.8\,{\scriptsize(1.5)} & 39.3\,{\scriptsize(2.1)} & 0.76\,{\scriptsize(0.01)} & \multicolumn{1}{c}{--} & 0.00\,{\scriptsize(0.00)} & 25.9\,{\scriptsize(1.2)} \\
    FS-PC & 70.9\,{\scriptsize(0.6)} & 0.66\,{\scriptsize(0.00)} & \multicolumn{1}{c}{--} & 0.00\,{\scriptsize(0.00)} & 39.7\,{\scriptsize(0.4)} & 84.1\,{\scriptsize(0.6)} & 0.60\,{\scriptsize(0.00)} & \multicolumn{1}{c}{--} & 0.00\,{\scriptsize(0.00)} & 51.9\,{\scriptsize(0.4)} \\
    FS-GES {\scriptsize(99/94)} & 168.3\,{\scriptsize(2.0)} & 0.37\,{\scriptsize(0.00)} & \multicolumn{1}{c}{--} & 0.00\,{\scriptsize(0.00)} & 132.3\,{\scriptsize(1.9)} & 170.8\,{\scriptsize(1.7)} & 0.37\,{\scriptsize(0.00)} & \multicolumn{1}{c}{--} & 0.00\,{\scriptsize(0.00)} & 134.6\,{\scriptsize(1.7)} \\
    LC-GES & 57.8\,{\scriptsize(0.9)} & 0.73\,{\scriptsize(0.00)} & \multicolumn{1}{c}{--} & 0.00\,{\scriptsize(0.00)} & 29.8\,{\scriptsize(0.7)} & 50.2\,{\scriptsize(1.1)} & 0.77\,{\scriptsize(0.00)} & \multicolumn{1}{c}{--} & 0.00\,{\scriptsize(0.00)} & 24.0\,{\scriptsize(0.7)} \\
    \bottomrule
  \end{tabular}}
\end{table}

The comparison between DAGGLE and DAGGLE$^{(0)}$ illustrates the complementary roles of the two stages.
The continuous screening in Stage 1 captures the true adjacencies but retains spurious adjacencies, which are numerous on random and tree graphs. Consequently, on these graphs it recovers the true MEC much less often than the full estimator, whereas on star graphs the two rates are close.
Backward pruning removes most of these adjacencies on random and tree graphs and lowers the mean SHD in every setting.

The plug-in methods fail in the ways anticipated in Section~\ref{subsec:alternatives}.
The mean SHD of FS-PC increases with sample size across all six comparisons, as expected when factor scores alter the conditional independence relations and the method targets the wrong graph (Example~\ref{ex:proxy_ci_distortion} and Section~S1.1).
Similarly, FS-GES selects dense graphs, with more false-positive adjacencies than true edges in every setting, consistent with the additional nonzero entries in the precision matrix of regression scores (Section~S1.1).
LC-GES is the most accurate plug-in method on star and tree graphs and improves with $n$ in every setting, but it treats the estimated latent correlation matrix as a sample covariance matrix and thus understates its estimation uncertainty, which plausibly explains its many extra adjacencies.
In computation, DAGGLE finishes every replication within $8$ minutes at $m=20$ and within one hour at $m=40$, FS-PC and LC-GES finish within $6$ minutes, and FS-GES does not finish every tree replication at $m=40$ within the $8$-hour budget.

\section{Application to the European Social Survey}\label{sec:real}

\subsection{Data and Measurement Model}\label{subsec:ess_data}

The Trust in Justice module in Round~5 of the European Social Survey was developed to test the procedural justice theory of police legitimacy \citep{ess5data}.
According to this perspective, perceived procedural fairness and police effectiveness affect institutional legitimacy, which in turn affects public cooperation with the legal system \citep{tyler2006why,sunshine2003role,jackson2012why}.
Existing analyses of this survey rely on confirmatory structural equation models with path coefficients prescribed in advance by theoretical expectations \citep{jackson2013policing}.
We apply DAGGLE to explore three questions, namely what the indicator distribution alone determines about the latent graph, what a substantive background ordering resolves, and how the learned graph relates to the published structural equation model.

We analyze the Great Britain subsample comprising $n=1{,}805$ respondents with complete observations across fifteen survey items. Five latent constructs are measured, each by three items. 
Specifically, Effectiveness reflects perceived police success in preventing crime, apprehending burglars, and responding promptly. Procedural Fairness evaluates respectful treatment, impartial decision-making, and transparent communication by police officers. Obligation to Obey captures the felt moral duty to comply with police directives. Moral Alignment assesses the extent to which police officers are perceived as sharing citizens' ethical values and sense of justice. Finally, Cooperation measures willingness to report crimes, identify suspects, and provide evidence in court.
The Effectiveness and Obligation to Obey items are measured on an 11-point scale (0--10), the Procedural Fairness and Cooperation items on a four-point scale (1--4), and the Moral Alignment items on a five-point scale (1--5). All items are coded so that higher values reflect higher levels of the underlying construct, and are then standardized and treated as continuous indicators under the Gaussian likelihood.

Every indicator measures a unique construct, so $Q$ has simple structure with three indicators per construct and the measurement model is identified by the three-indicator rule of Remark~\ref{rem:measurement}. We apply DAGGLE using the standard penalty rate $\lambda_n=(\log n)/n$, the truncation levels of Section~\ref{subsec:sim_design}, and ten random initializations. With five constructs, the structural search space is small enough to enumerate, allowing us to evaluate the BIC of all $8{,}782$ MECs and every DAG compatible with the ordering introduced below. 

\subsection{Learned Structure and Markov Equivalence}\label{subsec:ess_results}

Table~\ref{tab:ess_candidates} evaluates four candidate structural models under common statistical criteria, namely the exhaustive BIC minimizer, the unrestricted and ordered DAGGLE selections, and the model published in \citet{jackson2013policing}  refitted as a DAG.
These models are compared by the number of DAGs in their Markov equivalence classes, their BIC values relative to the minimum, and likelihood-ratio tests against the saturated structural model.
In the absence of substantive ordering constraints, the data determine a set of core adjacencies but leave many edge directions undetermined.
Twenty candidate equivalence classes display nearly indistinguishable statistical fit, differing from the minimum BIC by less than the penalty for a single additional parameter. All twenty classes share the same seven core adjacencies, differing solely in edge orientations and in the inclusion of at most one of three peripheral adjacencies.
The unrestricted fit has a small equivalence class only because two weakly supported edges are absent, so its directions are not evidence from the data.

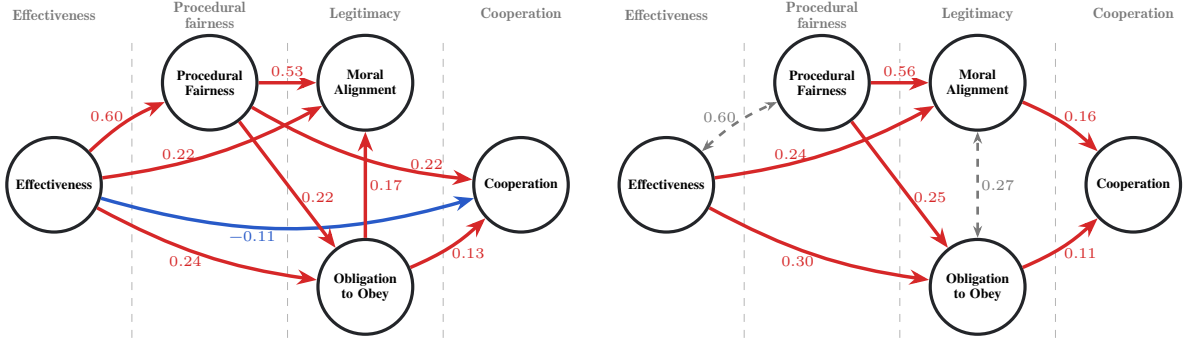
\begin{figure}[t]
  \centering
  \definecolor{essposred}{RGB}{214,39,40}%
  \definecolor{essnegblue}{RGB}{40,90,200}%
  \definecolor{essink}{RGB}{35,37,41}%
  \definecolor{essgray}{RGB}{120,120,120}%
  \begin{tikzpicture}[x=0.71cm,y=0.75cm,
      v/.style={circle, draw=essink, line width=1.2pt, fill=white, minimum size=12.5mm, text width=12mm,
          inner sep=0pt, align=center, font=\fontfamily{ptm}\fontseries{b}\fontsize{5.2}{5.7}\selectfont},
      pe/.style={-{Stealth[length=2.6mm,width=2mm]}, essposred, line width=1.3pt},
      ne/.style={-{Stealth[length=2.6mm,width=2mm]}, essnegblue, line width=1.3pt},
      cov/.style={{Stealth[length=1.6mm,width=1.2mm]}-{Stealth[length=1.6mm,width=1.2mm]}, essgray, dashed, line width=1pt},
      lab/.style={font=\fontfamily{lmr}\fontsize{5.6}{6}\selectfont, text=essposred, inner sep=1pt},
      labn/.style={font=\fontfamily{lmr}\fontsize{5.6}{6}\selectfont, text=essnegblue, inner sep=1pt},
      labg/.style={font=\fontfamily{lmr}\fontsize{5.6}{6}\selectfont, text=essgray, inner sep=1pt},
      tier/.style={font=\fontfamily{lmr}\fontsize{5.6}{6}\selectfont\bfseries, text=essgray, align=center},
      sep/.style={essgray!55, dashed, line width=0.35pt},
      ptitle/.style={font=\scriptsize, text=essink},
      lw/.style={line width=#1}]
  \node[ptitle] at (4.35,4.0) {(A) DAGGLE under the procedural-justice ordering};
  \node[tier] at (0,3.0)   {Effectiveness};
  \node[tier] at (2.9,3.0) {Procedural\\fairness};
  \node[tier] at (5.8,3.0) {Legitimacy};
  \node[tier] at (8.7,3.0) {Cooperation};
  \foreach \x in {1.45,4.35,7.25} \draw[sep] (\x,-2.7)--(\x,2.55);
  \node[v] (E) at (0,0)      {Effectiveness};
  \node[v] (F) at (2.9,1.8)  {Procedural\\Fairness};
  \node[v] (M) at (5.8,1.8)  {Moral\\Alignment};
  \node[v] (O) at (5.8,-1.8) {Obligation\\to Obey};
  \node[v] (C) at (8.7,0)    {Cooperation};
  \draw[pe] (E) to[bend left=10]  node[lab,above left=-1.5pt,pos=.5]{0.60} (F);
  \draw[pe] (E) to[bend right=10] node[lab,below left=-1.5pt,pos=.5]{0.24} (O);
  \draw[pe](E) to[bend right=10] node[lab,above,pos=.32]{0.22} (M);
  \draw[ne](E) to[bend right=16] node[labn,below,pos=.4]{\textminus0.11} (C);
  \draw[pe](F) -- node[lab,right,pos=.6]{0.22} (O);
  \draw[pe](F) -- node[lab,above=1pt,fill=white]{0.53} (M);
  \draw[pe](F) to[bend right=12] node[lab,above right=-1.5pt,pos=.74]{0.22} (C);
  \draw[pe](O) -- node[lab,right]{0.17} (M);
  \draw[pe] (O) to[bend right=10] node[lab,below right=-1.5pt,pos=.5]{0.13} (C);
  \begin{scope}[xshift=8.1cm]
  \node[ptitle] at (4.35,4.0) {(B) Structural equation model of \citet{jackson2013policing}};
  \node[tier] at (0,3.0)   {Effectiveness};
  \node[tier] at (2.9,3.0) {Procedural\\fairness};
  \node[tier] at (5.8,3.0) {Legitimacy};
  \node[tier] at (8.7,3.0) {Cooperation};
  \foreach \x in {1.45,4.35,7.25} \draw[sep] (\x,-2.7)--(\x,2.55);
  \node[v] (E) at (0,0)      {Effectiveness};
  \node[v] (F) at (2.9,1.8)  {Procedural\\Fairness};
  \node[v] (M) at (5.8,1.8)  {Moral\\Alignment};
  \node[v] (O) at (5.8,-1.8) {Obligation\\to Obey};
  \node[v] (C) at (8.7,0)    {Cooperation};
  \draw[cov] (E) to[bend left=10] node[labg,above left=-1.5pt,pos=.5]{0.60} (F);
  \draw[cov] (O) -- node[labg,right]{0.27} (M);
  \draw[pe] (E) to[bend right=10] node[lab,below left=-1.5pt,pos=.5]{0.30} (O);
  \draw[pe] (E) to[bend right=10] node[lab,above,pos=.32]{0.24} (M);
  \draw[pe] (F) -- node[lab,right,pos=.6]{0.25} (O);
  \draw[pe] (F) -- node[lab,above=1pt,fill=white]{0.56} (M);
  \draw[pe](O) to[bend right=10] node[lab,below right=-1.5pt,pos=.5]{0.11} (C);
  \draw[pe](M) to[bend left=10]  node[lab,above right=-1.5pt,pos=.5]{0.16} (C);
  \end{scope}
\end{tikzpicture}
  \caption{European Social Survey data. (A) The DAG selected by DAGGLE under the procedural-justice ordering. Its CPDAG is undirected, so the ordering determines every direction shown. (B) The published structural equation model on the same constructs. Constructs are arranged by tier from left to right. Directed edges are labeled with standardized path coefficients (red for positive and blue for negative), and grey dashed edges with covariance values.}
  \label{fig:ess_comparison}
\end{figure}

We therefore incorporate the theory-driven procedural justice ordering, arranging the constructs in the sequence of Effectiveness, Procedural Fairness, Obligation to Obey, Moral Alignment, and Cooperation.
The structural constraint matrix $\bZ$ permits all forward edges while ruling out reverse edges.
Under this specification, DAGGLE recovers the nine-edge DAG shown in Figure~\ref{fig:ess_comparison}(A), which achieves the global BIC minimum among all $1{,}024$ graphs compatible with the ordering.
Its BIC exceeds the smallest BIC over all MECs by $9.7$, somewhat more than the penalty $\log n\approx7.5$ for one parameter. This difference is the cost of imposing the ordering. 
Crucially, this ordered graph exhibits no detectable likelihood loss relative to the saturated structural model, confirming that omitting the edge from Moral Alignment to Cooperation does not compromise empirical fit. In contrast, the published structural equation model refitted to these data is rejected by the same likelihood ratio test.

\begin{table}[t]
  \centering
  \small
  \singlespacing
  \caption{Candidate structures for the five-construct analysis. Edges are counted without direction, and DAGs in MEC is the size of the Markov equivalence class. $\Delta$BIC is measured from the smallest BIC over all $8{,}782$ MECs. Likelihood loss is the likelihood-ratio statistic against the saturated structural model with all ten edges, with degrees of freedom in parentheses and the $p$-value of the chi-squared test. The published model is that of \citet{jackson2013policing}, written as a DAG under the ordering, excluding its nonsignificant path from Effectiveness to Cooperation, and refitted on these data.}
  \label{tab:ess_candidates}
  \setlength{\tabcolsep}{4pt}
  \resizebox{\linewidth}{!}{\begin{tabular}{l c c r r r}
    \toprule
    Structure & Edges & DAGs in MEC & $\Delta$BIC & Likelihood loss (df) & $p$ \\
    \midrule
    Exhaustive minimum over all MECs & 7 & 8  & 0.0  & 5.3 (3)  & 0.15 \\
    Unrestricted fit (DAGGLE)        & 8 & 3  & 2.3  & 0.1 (2)  & 0.95 \\
    Ordered fit (DAGGLE)             & 9 & 42 & 9.7  & 0.0 (1)  & 0.95 \\
    Published model as a DAG         & 8 & 32 & 23.7 & 21.6 (2) & $<0.001$ \\
    \bottomrule
  \end{tabular}}
\end{table}

Figure~\ref{fig:ess_comparison}(A) presents the ordered network structured across substantive tiers. Its nine edges comprise the seven core edges alongside Effectiveness--Cooperation and Procedural Fairness--Obligation to Obey. 
Among the $42$ observationally equivalent orientations belonging to this Markov equivalence class, the substantive ordering constraint uniquely identifies the DAG shown.
The data thus determine the seven core adjacencies, while the substantive ordering determines the edge directions and contributes to the selection of the two remaining adjacencies.

\subsection{Interpretation and Robustness}\label{subsec:ess_interp}

Figure~\ref{fig:ess_comparison}(B) shows the structural equation model of \citet{jackson2013policing} estimated on the same five constructs. 
The standardized coefficients on the directed paths shared by the two models are similar, including those from Procedural Fairness to Moral Alignment and from Obligation to Obey to Cooperation. The small negative coefficient on the path from Effectiveness to Cooperation also appears in both, although it is not significant in the published analysis and is not shown in panel~(B).
The agreement holds although the published analysis used the full United Kingdom sample and conditioned on additional background covariates.

Discrepancies between the two models occur where the data provide little information about the structure.
In the published model, edges between Effectiveness and Procedural Fairness and between Obligation to Obey and Moral Alignment are modeled as covariances without direction, whereas the ordered fit directs them in accordance with substantive tiers.
Furthermore, the secondary direct path into Cooperation originates from Procedural Fairness in the ordered model rather than from Moral Alignment as posited in the published specification. The edge between Moral Alignment and Cooperation is one of the three peripheral adjacencies that the data leave undetermined. 
The fit comparison favors the ordered DAGGLE fit. We reject the refitted published model in Section~\ref{subsec:ess_results} because it omits the edges into Cooperation from Procedural Fairness and Effectiveness, the first of which is among the seven edges that all twenty near-best MECs contain.%

\begin{table}[t]
  \centering
  \small
  \singlespacing
  \caption{Standardized path coefficients of the ordered fit in Figure~\ref{fig:ess_comparison}(A) for the full Great Britain sample and for subgroups defined by sex, recent criminal victimization, and urbanicity. A dash marks an edge removed by BIC pruning in that subgroup.}
  \label{tab:ess_subgroups}
  \begin{tabular}{l rrrrrrr}
    \toprule
                         & Full    & Male    & Female  & Victim  & Non-victim & Urban   & Non-urban \\
    \cmidrule(lr){2-8}
    $n$                  & 1805    & 845     & 960     & 360     & 1445       & 576     & 1226      \\
    \midrule
    Eff $\to$ ProcFair   & $+0.60$ & $+0.57$ & $+0.64$ & $+0.70$ & $+0.55$    & $+0.61$ & $+0.59$   \\
    Eff $\to$ Obey       & $+0.24$ & $+0.22$ & $+0.27$ & $+0.41$ & $+0.22$    & $+0.24$ & $+0.23$   \\
    Eff $\to$ Moral      & $+0.22$ & $+0.31$ & --      & $+0.23$ & $+0.21$    & $+0.23$ & $+0.22$   \\
    Eff $\to$ Coop       & $-0.11$ & $-0.15$ & --      & --      & --         & --      & --        \\
    ProcFair $\to$ Obey  & $+0.22$ & $+0.27$ & $+0.17$ & --      & $+0.24$    & $+0.30$ & $+0.20$   \\
    ProcFair $\to$ Moral & $+0.53$ & $+0.45$ & $+0.70$ & $+0.51$ & $+0.54$    & $+0.65$ & $+0.49$   \\
    ProcFair $\to$ Coop  & $+0.22$ & $+0.27$ & $+0.17$ & $+0.19$ & $+0.13$    & $+0.20$ & $+0.14$   \\
    Obey $\to$ Moral     & $+0.17$ & $+0.20$ & $+0.18$ & $+0.20$ & $+0.17$    & --      & $+0.22$   \\
    Obey $\to$ Coop      & $+0.13$ & $+0.17$ & --      & $+0.20$ & $+0.09$    & $+0.14$ & --        \\
    \bottomrule
  \end{tabular}
\end{table}

The selected structure is stable. Table~\ref{tab:ess_subgroups} refits the ordered model across demographic subgroups stratified by respondent sex, recent criminal victimization, and urbanicity. Every retained edge exhibits a consistent sign across all subgroups. The edges with the largest coefficients are retained in all subgroups, and only edges with small coefficients are removed in some subgroups. This pattern is expected, because the likelihood gain from an edge is proportional to the sample size while the BIC penalty grows only with its logarithm. Adding Fear of Crime as a sixth latent construct ordered prior to Effectiveness leaves the nine edges and their coefficients essentially unchanged (Section~S4). 

\section{Discussion}\label{sec:dis} 

In this article, we introduced a semi-exploratory framework for learning structural DAGs in Gaussian linear SEMs under a prespecified measurement structure.
By separating measurement identification from structural graph discovery, we establish that the structural DAG is identifiable up to its Markov equivalence class under standard identification and faithfulness assumptions. To learn this structure, the proposed DAGGLE procedure combines continuous screening with BIC-guided backward pruning, estimating measurement and structural parameters jointly through the marginal likelihood of the observed indicators. Its asymptotic consistency in fixed-dimensional settings follows from the argument of \citet{ZhangChenMoustaki2026} under standard likelihood regularity, numerical optimization precision, and backward search regularity. Our simulations demonstrate the substantial benefits of full-information joint estimation over proxy factor-score methods and highlight the role of backward pruning in removing spurious edges. Furthermore, our application to the European Social Survey data illustrates how DAGGLE can complement confirmatory SEM by uncovering structural relationships without requiring a fully specified path diagram. Finally, although our simulations and empirical application focus on pure measurement models, both DAGGLE and its theoretical guarantees readily accommodate prespecified cross-loadings under the assumptions of Section~\ref{sec:iden}.

The learned structure requires care in causal interpretation, as with any structural equation model. Under the observational Gaussian model, the data generally identify a MEC rather than a unique DAG. An edge direction chosen for one representative should therefore not be interpreted as empirically established when it is reversible within the class. Even compelled edges that are shared across all members of an equivalence class do not automatically establish causality. A causal interpretation also depends on the validity of the measurement specification, the assumption of independent structural disturbances, and broader substantive considerations. Prior knowledge can resolve otherwise unidentified directions, but its contribution should be distinguished from the information supplied by the data. In the ESS application, for example, every direction in the ordered fit is determined by the imposed substantive ordering. The learned graph thus provides structured, data-consistent hypotheses for subsequent empirical investigation rather than empirical proof of the causal ordering.

Several extensions remain for future investigation. First, alternative smooth characterizations of acyclicity, such as the log-determinant formulation of \cite{bello2022dagma}, may improve optimization performance in the screening stage. 
Second, combining observational data with experimental interventions or longitudinal panel data could help resolve observationally equivalent graphs within an MEC, potentially identifying a unique DAG under suitable assumptions.
Third, further studies could assess finite-sample performance with prespecified cross-loadings and sensitivity to measurement misspecifications, such as unmodeled cross-loadings and residual local dependencies. The ESS findings should also be interpreted in light of the continuous approximation to ordinal responses. Extensions to nonlinear structural relationships and
high-dimensional regimes where the number of constructs grows with sample size would further broaden the scope of the framework.

\bibliographystyle{apacite-noinit}
\bibliography{ref}

\begin{thebibliography}{}

\bibitem [\protect \citeauthoryear {%
Anderson%
\ \BBA {} Rubin%
}{%
Anderson%
\ \BBA {} Rubin%
}{%
{\protect \APACyear {1956}}%
}]{%
anderson1956statistical}
\APACinsertmetastar {%
anderson1956statistical}%
\begin{APACrefauthors}%
Anderson, T\BPBI W.%
\BCBT {}\ \BBA {} Rubin, H.%
\end{APACrefauthors}%
\unskip\
\newblock
\APACrefYearMonthDay{1956}{}{}.
\newblock
{\BBOQ}\APACrefatitle {Statistical Inference in Factor Analysis} {Statistical
  inference in factor analysis}.{\BBCQ}
\newblock
\BIn{} J.~Neyman\ (\BED), \APACrefbtitle {Proceedings of the Third {Berkeley}
  Symposium on Mathematical Statistics and Probability} {Proceedings of the
  third {Berkeley} symposium on mathematical statistics and probability}\
  (\BVOL~5, \BPGS\ 111--150).
\newblock
\APACaddressPublisher{Berkeley}{University of California Press}.
\PrintBackRefs{\CurrentBib}

\bibitem [\protect \citeauthoryear {%
Andersson%
, Madigan%
\BCBL {}\ \BBA {} Perlman%
}{%
Andersson%
\ \protect \BOthers {.}}{%
{\protect \APACyear {1997}}%
{\protect \APACexlab {{\protect \BCnt {1}}}}}]{%
andersson1997characterization}
\APACinsertmetastar {%
andersson1997characterization}%
\begin{APACrefauthors}%
Andersson, S\BPBI A.%
, Madigan, D.%
\BCBL {}\ \BBA {} Perlman, M\BPBI D.%
\end{APACrefauthors}%
\unskip\
\newblock
\APACrefYearMonthDay{1997{\protect \BCnt {1}}}{}{}.
\newblock
{\BBOQ}\APACrefatitle {A Characterization of {M}arkov Equivalence Classes for
  Acyclic Digraphs} {A characterization of {M}arkov equivalence classes for
  acyclic digraphs}.{\BBCQ}
\newblock
\APACjournalVolNumPages{The Annals of Statistics}{25}{2}{505--541}.
\PrintBackRefs{\CurrentBib}

\bibitem [\protect \citeauthoryear {%
Andersson%
, Madigan%
\BCBL {}\ \BBA {} Perlman%
}{%
Andersson%
\ \protect \BOthers {.}}{%
{\protect \APACyear {1997}}%
{\protect \APACexlab {{\protect \BCnt {2}}}}}]{%
AnderssonMadiganPerlman1997}
\APACinsertmetastar {%
AnderssonMadiganPerlman1997}%
\begin{APACrefauthors}%
Andersson, S\BPBI A.%
, Madigan, D.%
\BCBL {}\ \BBA {} Perlman, M\BPBI D.%
\end{APACrefauthors}%
\unskip\
\newblock
\APACrefYearMonthDay{1997{\protect \BCnt {2}}}{}{}.
\newblock
{\BBOQ}\APACrefatitle {On the {M}arkov Equivalence of Chain Graphs, Undirected
  Graphs, and Acyclic Digraphs} {On the {M}arkov equivalence of chain graphs,
  undirected graphs, and acyclic digraphs}.{\BBCQ}
\newblock
\APACjournalVolNumPages{Scandinavian Journal of Statistics}{24}{1}{81--102}.
\PrintBackRefs{\CurrentBib}

\bibitem [\protect \citeauthoryear {%
Aragam%
\ \BBA {} Zhou%
}{%
Aragam%
\ \BBA {} Zhou%
}{%
{\protect \APACyear {2015}}%
}]{%
aragam2015concave}
\APACinsertmetastar {%
aragam2015concave}%
\begin{APACrefauthors}%
Aragam, B.%
\BCBT {}\ \BBA {} Zhou, Q.%
\end{APACrefauthors}%
\unskip\
\newblock
\APACrefYearMonthDay{2015}{}{}.
\newblock
{\BBOQ}\APACrefatitle {Concave Penalized Estimation of Sparse {G}aussian
  {B}ayesian Networks} {Concave penalized estimation of sparse {G}aussian
  {B}ayesian networks}.{\BBCQ}
\newblock
\APACjournalVolNumPages{Journal of Machine Learning
  Research}{16}{69}{2273--2328}.
\PrintBackRefs{\CurrentBib}

\bibitem [\protect \citeauthoryear {%
Asparouhov%
\ \BBA {} Muth{\'e}n%
}{%
Asparouhov%
\ \BBA {} Muth{\'e}n%
}{%
{\protect \APACyear {2009}}%
}]{%
AsparouhovMuthen2009}
\APACinsertmetastar {%
AsparouhovMuthen2009}%
\begin{APACrefauthors}%
Asparouhov, T.%
\BCBT {}\ \BBA {} Muth{\'e}n, B.%
\end{APACrefauthors}%
\unskip\
\newblock
\APACrefYearMonthDay{2009}{}{}.
\newblock
{\BBOQ}\APACrefatitle {Exploratory Structural Equation Modeling} {Exploratory
  structural equation modeling}.{\BBCQ}
\newblock
\APACjournalVolNumPages{Structural Equation Modeling: A Multidisciplinary
  Journal}{16}{3}{397--438}.
\PrintBackRefs{\CurrentBib}

\bibitem [\protect \citeauthoryear {%
Bartholomew%
, Knott%
\BCBL {}\ \BBA {} Moustaki%
}{%
Bartholomew%
\ \protect \BOthers {.}}{%
{\protect \APACyear {2011}}%
}]{%
bartholomew2011latent}
\APACinsertmetastar {%
bartholomew2011latent}%
\begin{APACrefauthors}%
Bartholomew, D\BPBI J.%
, Knott, M.%
\BCBL {}\ \BBA {} Moustaki, I.%
\end{APACrefauthors}%
\unskip\
\newblock
\APACrefYear{2011}.
\newblock
\APACrefbtitle {Latent Variable Models and Factor Analysis: A Unified Approach}
  {Latent variable models and factor analysis: A unified approach}\
  (\PrintOrdinal{3rd}\ \BEd).
\newblock
\APACaddressPublisher{Chichester}{John Wiley \& Sons}.
\PrintBackRefs{\CurrentBib}

\bibitem [\protect \citeauthoryear {%
Bello%
, Aragam%
\BCBL {}\ \BBA {} Ravikumar%
}{%
Bello%
\ \protect \BOthers {.}}{%
{\protect \APACyear {2022}}%
}]{%
bello2022dagma}
\APACinsertmetastar {%
bello2022dagma}%
\begin{APACrefauthors}%
Bello, K.%
, Aragam, B.%
\BCBL {}\ \BBA {} Ravikumar, P.%
\end{APACrefauthors}%
\unskip\
\newblock
\APACrefYearMonthDay{2022}{}{}.
\newblock
{\BBOQ}\APACrefatitle {{DAGMA}: Learning {DAGs} via {M}-Matrices and a
  Log-Determinant Acyclicity Characterization} {{DAGMA}: Learning {DAGs} via
  {M}-matrices and a log-determinant acyclicity characterization}.{\BBCQ}
\newblock
\BIn{} \APACrefbtitle {Advances in Neural Information Processing Systems}
  {Advances in neural information processing systems}\ (\BVOL~35, \BPGS\
  8226--8239).
\PrintBackRefs{\CurrentBib}

\bibitem [\protect \citeauthoryear {%
Bollen%
}{%
Bollen%
}{%
{\protect \APACyear {1989}}%
}]{%
bollen1989structural}
\APACinsertmetastar {%
bollen1989structural}%
\begin{APACrefauthors}%
Bollen, K\BPBI A.%
\end{APACrefauthors}%
\unskip\
\newblock
\APACrefYear{1989}.
\newblock
\APACrefbtitle {Structural Equations with Latent Variables} {Structural
  equations with latent variables}.
\newblock
\APACaddressPublisher{New York}{Wiley}.
\PrintBackRefs{\CurrentBib}

\bibitem [\protect \citeauthoryear {%
Bollen%
\ \BBA {} Pearl%
}{%
Bollen%
\ \BBA {} Pearl%
}{%
{\protect \APACyear {2013}}%
}]{%
BollenPearl2013}
\APACinsertmetastar {%
BollenPearl2013}%
\begin{APACrefauthors}%
Bollen, K\BPBI A.%
\BCBT {}\ \BBA {} Pearl, J.%
\end{APACrefauthors}%
\unskip\
\newblock
\APACrefYearMonthDay{2013}{}{}.
\newblock
{\BBOQ}\APACrefatitle {Eight Myths About Causality and Structural Equation
  Models} {Eight myths about causality and structural equation models}.{\BBCQ}
\newblock
\BIn{} S\BPBI L.~Morgan\ (\BED), \APACrefbtitle {Handbook of Causal Analysis
  for Social Research} {Handbook of causal analysis for social research}\
  (\BPGS\ 301--328).
\newblock
\APACaddressPublisher{Dordrecht}{Springer}.
\PrintBackRefs{\CurrentBib}

\bibitem [\protect \citeauthoryear {%
Borsboom%
}{%
Borsboom%
}{%
{\protect \APACyear {2022}}%
}]{%
Borsboom2022}
\APACinsertmetastar {%
Borsboom2022}%
\begin{APACrefauthors}%
Borsboom, D.%
\end{APACrefauthors}%
\unskip\
\newblock
\APACrefYearMonthDay{2022}{}{}.
\newblock
{\BBOQ}\APACrefatitle {Possible Futures for Network Psychometrics} {Possible
  futures for network psychometrics}.{\BBCQ}
\newblock
\APACjournalVolNumPages{Psychometrika}{87}{1}{253--265}.
\PrintBackRefs{\CurrentBib}

\bibitem [\protect \citeauthoryear {%
Borsboom%
\ \BBA {} Cramer%
}{%
Borsboom%
\ \BBA {} Cramer%
}{%
{\protect \APACyear {2013}}%
}]{%
BorsboomCramer2013}
\APACinsertmetastar {%
BorsboomCramer2013}%
\begin{APACrefauthors}%
Borsboom, D.%
\BCBT {}\ \BBA {} Cramer, A\BPBI O\BPBI J.%
\end{APACrefauthors}%
\unskip\
\newblock
\APACrefYearMonthDay{2013}{}{}.
\newblock
{\BBOQ}\APACrefatitle {Network Analysis: An Integrative Approach to the
  Structure of Psychopathology} {Network analysis: An integrative approach to
  the structure of psychopathology}.{\BBCQ}
\newblock
\APACjournalVolNumPages{Annual Review of Clinical Psychology}{9}{}{91--121}.
\PrintBackRefs{\CurrentBib}

\bibitem [\protect \citeauthoryear {%
Borsboom%
\ \protect \BOthers {.}}{%
Borsboom%
\ \protect \BOthers {.}}{%
{\protect \APACyear {2021}}%
}]{%
BorsboomEtAl2021}
\APACinsertmetastar {%
BorsboomEtAl2021}%
\begin{APACrefauthors}%
Borsboom, D.%
, Deserno, M\BPBI K.%
, Rhemtulla, M.%
, Epskamp, S.%
, Fried, E\BPBI I.%
, McNally, R\BPBI J.%
\BDBL {}Waldorp, L\BPBI J.%
\end{APACrefauthors}%
\unskip\
\newblock
\APACrefYearMonthDay{2021}{}{}.
\newblock
{\BBOQ}\APACrefatitle {Network Analysis of Multivariate Data in Psychological
  Science} {Network analysis of multivariate data in psychological
  science}.{\BBCQ}
\newblock
\APACjournalVolNumPages{Nature Reviews Methods Primers}{1}{}{58}.
\PrintBackRefs{\CurrentBib}

\bibitem [\protect \citeauthoryear {%
Castelletti%
}{%
Castelletti%
}{%
{\protect \APACyear {2024}}%
}]{%
Castelletti2024}
\APACinsertmetastar {%
Castelletti2024}%
\begin{APACrefauthors}%
Castelletti, F.%
\end{APACrefauthors}%
\unskip\
\newblock
\APACrefYearMonthDay{2024}{}{}.
\newblock
{\BBOQ}\APACrefatitle {Learning {Bayesian} Networks: A Copula Approach for
  Mixed-Type Data} {Learning {Bayesian} networks: A copula approach for
  mixed-type data}.{\BBCQ}
\newblock
\APACjournalVolNumPages{Psychometrika}{89}{2}{658--686}.
\PrintBackRefs{\CurrentBib}

\bibitem [\protect \citeauthoryear {%
Chickering%
}{%
Chickering%
}{%
{\protect \APACyear {2002}}%
{\protect \APACexlab {{\protect \BCnt {1}}}}}]{%
chickering2002equivalence}
\APACinsertmetastar {%
chickering2002equivalence}%
\begin{APACrefauthors}%
Chickering, D\BPBI M.%
\end{APACrefauthors}%
\unskip\
\newblock
\APACrefYearMonthDay{2002{\protect \BCnt {1}}}{}{}.
\newblock
{\BBOQ}\APACrefatitle {Learning Equivalence Classes of {B}ayesian-Network
  Structures} {Learning equivalence classes of {B}ayesian-network
  structures}.{\BBCQ}
\newblock
\APACjournalVolNumPages{Journal of Machine Learning Research}{2}{}{445--498}.
\PrintBackRefs{\CurrentBib}

\bibitem [\protect \citeauthoryear {%
Chickering%
}{%
Chickering%
}{%
{\protect \APACyear {2002}}%
{\protect \APACexlab {{\protect \BCnt {2}}}}}]{%
chickering2002optimal}
\APACinsertmetastar {%
chickering2002optimal}%
\begin{APACrefauthors}%
Chickering, D\BPBI M.%
\end{APACrefauthors}%
\unskip\
\newblock
\APACrefYearMonthDay{2002{\protect \BCnt {2}}}{}{}.
\newblock
{\BBOQ}\APACrefatitle {Optimal Structure Identification with Greedy Search}
  {Optimal structure identification with greedy search}.{\BBCQ}
\newblock
\APACjournalVolNumPages{Journal of Machine Learning Research}{3}{}{507--554}.
\PrintBackRefs{\CurrentBib}

\bibitem [\protect \citeauthoryear {%
Croon%
}{%
Croon%
}{%
{\protect \APACyear {2002}}%
}]{%
Croon2002}
\APACinsertmetastar {%
Croon2002}%
\begin{APACrefauthors}%
Croon, M\BPBI A.%
\end{APACrefauthors}%
\unskip\
\newblock
\APACrefYearMonthDay{2002}{}{}.
\newblock
{\BBOQ}\APACrefatitle {Using Predicted Latent Scores in General Latent
  Structure Models} {Using predicted latent scores in general latent structure
  models}.{\BBCQ}
\newblock
\BIn{} G\BPBI A.~Marcoulides\ \BBA {} I.~Moustaki\ (\BEDS), \APACrefbtitle
  {Latent Variable and Latent Structure Models} {Latent variable and latent
  structure models}\ (\BPGS\ 195--224).
\newblock
\APACaddressPublisher{Mahwah, NJ}{Lawrence Erlbaum Associates}.
\PrintBackRefs{\CurrentBib}

\bibitem [\protect \citeauthoryear {%
Dempster%
}{%
Dempster%
}{%
{\protect \APACyear {1972}}%
}]{%
Dempster1972}
\APACinsertmetastar {%
Dempster1972}%
\begin{APACrefauthors}%
Dempster, A\BPBI P.%
\end{APACrefauthors}%
\unskip\
\newblock
\APACrefYearMonthDay{1972}{}{}.
\newblock
{\BBOQ}\APACrefatitle {Covariance Selection} {Covariance selection}.{\BBCQ}
\newblock
\APACjournalVolNumPages{Biometrics}{28}{1}{157--175}.
\PrintBackRefs{\CurrentBib}

\bibitem [\protect \citeauthoryear {%
Devlieger%
, Mayer%
\BCBL {}\ \BBA {} Rosseel%
}{%
Devlieger%
\ \protect \BOthers {.}}{%
{\protect \APACyear {2016}}%
}]{%
DevliegerMayerRosseel2016}
\APACinsertmetastar {%
DevliegerMayerRosseel2016}%
\begin{APACrefauthors}%
Devlieger, I.%
, Mayer, A.%
\BCBL {}\ \BBA {} Rosseel, Y.%
\end{APACrefauthors}%
\unskip\
\newblock
\APACrefYearMonthDay{2016}{}{}.
\newblock
{\BBOQ}\APACrefatitle {Hypothesis Testing Using Factor Score Regression: A
  Comparison of Four Methods} {Hypothesis testing using factor score
  regression: A comparison of four methods}.{\BBCQ}
\newblock
\APACjournalVolNumPages{Educational and Psychological
  Measurement}{76}{5}{741--770}.
\PrintBackRefs{\CurrentBib}

\bibitem [\protect \citeauthoryear {%
Dong%
\ \protect \BOthers {.}}{%
Dong%
\ \protect \BOthers {.}}{%
{\protect \APACyear {2026}}%
}]{%
DongEtAl2026LGES}
\APACinsertmetastar {%
DongEtAl2026LGES}%
\begin{APACrefauthors}%
Dong, X.%
, Ng, I.%
, Dai, H.%
, Sun, J.%
, Song, X.%
, Spirtes, P.%
\BCBL {}\ \BBA {} Zhang, K.%
\end{APACrefauthors}%
\unskip\
\newblock
\APACrefYearMonthDay{2026}{}{}.
\newblock
{\BBOQ}\APACrefatitle {Score-Based Greedy Search for Structure Identification
  of Partially Observed Causal Models} {Score-based greedy search for structure
  identification of partially observed causal models}.{\BBCQ}
\newblock
\BIn{} \APACrefbtitle {International Conference on Learning Representations.}
  {International conference on learning representations.}
\PrintBackRefs{\CurrentBib}

\bibitem [\protect \citeauthoryear {%
Epskamp%
, Borsboom%
\BCBL {}\ \BBA {} Fried%
}{%
Epskamp%
, Borsboom%
\BCBL {}\ \BBA {} Fried%
}{%
{\protect \APACyear {2018}}%
}]{%
EpskampBorsboomFried2018}
\APACinsertmetastar {%
EpskampBorsboomFried2018}%
\begin{APACrefauthors}%
Epskamp, S.%
, Borsboom, D.%
\BCBL {}\ \BBA {} Fried, E\BPBI I.%
\end{APACrefauthors}%
\unskip\
\newblock
\APACrefYearMonthDay{2018}{}{}.
\newblock
{\BBOQ}\APACrefatitle {Estimating Psychological Networks and Their Accuracy: A
  Tutorial Paper} {Estimating psychological networks and their accuracy: A
  tutorial paper}.{\BBCQ}
\newblock
\APACjournalVolNumPages{Behavior Research Methods}{50}{1}{195--212}.
\PrintBackRefs{\CurrentBib}

\bibitem [\protect \citeauthoryear {%
Epskamp%
, Rhemtulla%
\BCBL {}\ \BBA {} Borsboom%
}{%
Epskamp%
\ \protect \BOthers {.}}{%
{\protect \APACyear {2017}}%
}]{%
EpskampRhemtullaBorsboom2017}
\APACinsertmetastar {%
EpskampRhemtullaBorsboom2017}%
\begin{APACrefauthors}%
Epskamp, S.%
, Rhemtulla, M.%
\BCBL {}\ \BBA {} Borsboom, D.%
\end{APACrefauthors}%
\unskip\
\newblock
\APACrefYearMonthDay{2017}{}{}.
\newblock
{\BBOQ}\APACrefatitle {Generalized Network Psychometrics: Combining Network and
  Latent Variable Models} {Generalized network psychometrics: Combining network
  and latent variable models}.{\BBCQ}
\newblock
\APACjournalVolNumPages{Psychometrika}{82}{4}{904--927}.
\PrintBackRefs{\CurrentBib}

\bibitem [\protect \citeauthoryear {%
Epskamp%
, Waldorp%
, M{\~o}ttus%
\BCBL {}\ \BBA {} Borsboom%
}{%
Epskamp%
, Waldorp%
\BCBL {}\ \protect \BOthers {.}}{%
{\protect \APACyear {2018}}%
}]{%
EpskampWaldorpMottusBorsboom2018}
\APACinsertmetastar {%
EpskampWaldorpMottusBorsboom2018}%
\begin{APACrefauthors}%
Epskamp, S.%
, Waldorp, L\BPBI J.%
, M{\~o}ttus, R.%
\BCBL {}\ \BBA {} Borsboom, D.%
\end{APACrefauthors}%
\unskip\
\newblock
\APACrefYearMonthDay{2018}{}{}.
\newblock
{\BBOQ}\APACrefatitle {The {G}aussian Graphical Model in Cross-Sectional and
  Time-Series Data} {The {G}aussian graphical model in cross-sectional and
  time-series data}.{\BBCQ}
\newblock
\APACjournalVolNumPages{Multivariate Behavioral Research}{53}{4}{453--480}.
\PrintBackRefs{\CurrentBib}

\bibitem [\protect \citeauthoryear {%
{European Social Survey European Research Infrastructure}%
}{%
{European Social Survey European Research Infrastructure}%
}{%
{\protect \APACyear {2025}}%
}]{%
ess5data}
\APACinsertmetastar {%
ess5data}%
\begin{APACrefauthors}%
{European Social Survey European Research Infrastructure}.%
\end{APACrefauthors}%
\unskip\
\newblock
\APACrefYearMonthDay{2025}{}{}.
\newblock
\APACrefbtitle {{ESS5} -- Integrated File, Edition 3.6 ({A}ustria Not
  Included)} {{ESS5} -- integrated file, edition 3.6 ({A}ustria not included)}\
  [Data set].
\newblock
\APAChowpublished {Sikt -- Norwegian Agency for Shared Services in Education
  and Research}.
\newblock
\begin{APACrefDOI} \doi{10.21338/ess5e03_6} \end{APACrefDOI}
\PrintBackRefs{\CurrentBib}

\bibitem [\protect \citeauthoryear {%
Gu%
\ \BBA {} Xu%
}{%
Gu%
\ \BBA {} Xu%
}{%
{\protect \APACyear {2019}}%
}]{%
GuXu2019}
\APACinsertmetastar {%
GuXu2019}%
\begin{APACrefauthors}%
Gu, Y.%
\BCBT {}\ \BBA {} Xu, G.%
\end{APACrefauthors}%
\unskip\
\newblock
\APACrefYearMonthDay{2019}{}{}.
\newblock
{\BBOQ}\APACrefatitle {Learning Attribute Patterns in High-Dimensional
  Structured Latent Attribute Models} {Learning attribute patterns in
  high-dimensional structured latent attribute models}.{\BBCQ}
\newblock
\APACjournalVolNumPages{Journal of Machine Learning Research}{20}{115}{1--58}.
\PrintBackRefs{\CurrentBib}

\bibitem [\protect \citeauthoryear {%
Huang%
, Chen%
\BCBL {}\ \BBA {} Weng%
}{%
Huang%
\ \protect \BOthers {.}}{%
{\protect \APACyear {2017}}%
}]{%
HuangChenWeng2017}
\APACinsertmetastar {%
HuangChenWeng2017}%
\begin{APACrefauthors}%
Huang, P\BHBI H.%
, Chen, H.%
\BCBL {}\ \BBA {} Weng, L\BHBI J.%
\end{APACrefauthors}%
\unskip\
\newblock
\APACrefYearMonthDay{2017}{}{}.
\newblock
{\BBOQ}\APACrefatitle {A Penalized Likelihood Method for Structural Equation
  Modeling} {A penalized likelihood method for structural equation
  modeling}.{\BBCQ}
\newblock
\APACjournalVolNumPages{Psychometrika}{82}{2}{329--354}.
\PrintBackRefs{\CurrentBib}

\bibitem [\protect \citeauthoryear {%
Jackson%
, Bradford%
\BCBL {}\ \protect \BOthers {.}}{%
Jackson%
, Bradford%
\BCBL {}\ \protect \BOthers {.}}{%
{\protect \APACyear {2012}}%
}]{%
jackson2012why}
\APACinsertmetastar {%
jackson2012why}%
\begin{APACrefauthors}%
Jackson, J.%
, Bradford, B.%
, Hough, M.%
, Myhill, A.%
, Quinton, P.%
\BCBL {}\ \BBA {} Tyler, T\BPBI R.%
\end{APACrefauthors}%
\unskip\
\newblock
\APACrefYearMonthDay{2012}{}{}.
\newblock
{\BBOQ}\APACrefatitle {Why Do People Comply with the Law? {Legitimacy} and the
  Influence of Legal Institutions} {Why do people comply with the law?
  {Legitimacy} and the influence of legal institutions}.{\BBCQ}
\newblock
\APACjournalVolNumPages{British Journal of Criminology}{52}{6}{1051--1071}.
\PrintBackRefs{\CurrentBib}

\bibitem [\protect \citeauthoryear {%
Jackson%
, Hough%
, Bradford%
, Hohl%
\BCBL {}\ \BBA {} Kuha%
}{%
Jackson%
, Hough%
\BCBL {}\ \protect \BOthers {.}}{%
{\protect \APACyear {2012}}%
}]{%
jackson2013policing}
\APACinsertmetastar {%
jackson2013policing}%
\begin{APACrefauthors}%
Jackson, J.%
, Hough, M.%
, Bradford, B.%
, Hohl, K.%
\BCBL {}\ \BBA {} Kuha, J.%
\end{APACrefauthors}%
\unskip\
\newblock
\APACrefYearMonthDay{2012}{}{}.
\newblock
\APACrefbtitle {Policing by Consent: Understanding the Dynamics of Police Power
  and Legitimacy} {Policing by consent: Understanding the dynamics of police
  power and legitimacy}\ \APACbVolEdTR {}{{ESS} Country Specific Topline
  Results Series\ \BNUM~1}.
\newblock
\APACaddressInstitution{London}{European Social Survey}.
\newblock
\APACrefnote{Trust in Justice module, {ESS} Round 5}
\PrintBackRefs{\CurrentBib}

\bibitem [\protect \citeauthoryear {%
Jacobucci%
, Grimm%
\BCBL {}\ \BBA {} McArdle%
}{%
Jacobucci%
\ \protect \BOthers {.}}{%
{\protect \APACyear {2016}}%
}]{%
JacobucciGrimmMcArdle2016}
\APACinsertmetastar {%
JacobucciGrimmMcArdle2016}%
\begin{APACrefauthors}%
Jacobucci, R.%
, Grimm, K\BPBI J.%
\BCBL {}\ \BBA {} McArdle, J\BPBI J.%
\end{APACrefauthors}%
\unskip\
\newblock
\APACrefYearMonthDay{2016}{}{}.
\newblock
{\BBOQ}\APACrefatitle {Regularized Structural Equation Modeling} {Regularized
  structural equation modeling}.{\BBCQ}
\newblock
\APACjournalVolNumPages{Structural Equation Modeling: A Multidisciplinary
  Journal}{23}{4}{555--566}.
\PrintBackRefs{\CurrentBib}

\bibitem [\protect \citeauthoryear {%
J\"{o}reskog%
}{%
J\"{o}reskog%
}{%
{\protect \APACyear {1969}}%
}]{%
joreskog:69}
\APACinsertmetastar {%
joreskog:69}%
\begin{APACrefauthors}%
J\"{o}reskog, K\BPBI G.%
\end{APACrefauthors}%
\unskip\
\newblock
\APACrefYearMonthDay{1969}{}{}.
\newblock
{\BBOQ}\APACrefatitle {A general approach to confirmatory maximum likelihood
  factor analysis} {A general approach to confirmatory maximum likelihood
  factor analysis}.{\BBCQ}
\newblock
\APACjournalVolNumPages{Psychometrika}{34}{2}{183--202}.
\PrintBackRefs{\CurrentBib}

\bibitem [\protect \citeauthoryear {%
J{\"o}reskog%
}{%
J{\"o}reskog%
}{%
{\protect \APACyear {1973}}%
}]{%
Joreskog1973}
\APACinsertmetastar {%
Joreskog1973}%
\begin{APACrefauthors}%
J{\"o}reskog, K\BPBI G.%
\end{APACrefauthors}%
\unskip\
\newblock
\APACrefYearMonthDay{1973}{}{}.
\newblock
{\BBOQ}\APACrefatitle {A General Method for Estimating a Linear Structural
  Equation System} {A general method for estimating a linear structural
  equation system}.{\BBCQ}
\newblock
\BIn{} A\BPBI S.~Goldberger\ \BBA {} O\BPBI D.~Duncan\ (\BEDS), \APACrefbtitle
  {Structural Equation Models in the Social Sciences} {Structural equation
  models in the social sciences}\ (\BPGS\ 85--112).
\newblock
\APACaddressPublisher{New York}{Seminar Press}.
\PrintBackRefs{\CurrentBib}

\bibitem [\protect \citeauthoryear {%
Koller%
\ \BBA {} Friedman%
}{%
Koller%
\ \BBA {} Friedman%
}{%
{\protect \APACyear {2009}}%
}]{%
KollerFriedman2009}
\APACinsertmetastar {%
KollerFriedman2009}%
\begin{APACrefauthors}%
Koller, D.%
\BCBT {}\ \BBA {} Friedman, N.%
\end{APACrefauthors}%
\unskip\
\newblock
\APACrefYear{2009}.
\newblock
\APACrefbtitle {Probabilistic Graphical Models: Principles and Techniques}
  {Probabilistic graphical models: Principles and techniques}.
\newblock
\APACaddressPublisher{Cambridge, MA}{MIT Press}.
\PrintBackRefs{\CurrentBib}

\bibitem [\protect \citeauthoryear {%
Lauritzen%
}{%
Lauritzen%
}{%
{\protect \APACyear {1996}}%
}]{%
lauritzen1996graphical}
\APACinsertmetastar {%
lauritzen1996graphical}%
\begin{APACrefauthors}%
Lauritzen, S\BPBI L.%
\end{APACrefauthors}%
\unskip\
\newblock
\APACrefYear{1996}.
\newblock
\APACrefbtitle {Graphical Models} {Graphical models}.
\newblock
\APACaddressPublisher{Oxford}{Oxford University Press}.
\PrintBackRefs{\CurrentBib}

\bibitem [\protect \citeauthoryear {%
Lee%
\ \BBA {} Gu%
}{%
Lee%
\ \BBA {} Gu%
}{%
{\protect \APACyear {2024}}%
}]{%
LeeGu2024}
\APACinsertmetastar {%
LeeGu2024}%
\begin{APACrefauthors}%
Lee, S.%
\BCBT {}\ \BBA {} Gu, Y.%
\end{APACrefauthors}%
\unskip\
\newblock
\APACrefYearMonthDay{2024}{}{}.
\newblock
{\BBOQ}\APACrefatitle {Latent Conjunctive {Bayesian} Network: Unify Attribute
  Hierarchy and {Bayesian} Network for Cognitive Diagnosis} {Latent conjunctive
  {Bayesian} network: Unify attribute hierarchy and {Bayesian} network for
  cognitive diagnosis}.{\BBCQ}
\newblock
\APACjournalVolNumPages{The Annals of Applied Statistics}{18}{3}{1988--2011}.
\PrintBackRefs{\CurrentBib}

\bibitem [\protect \citeauthoryear {%
Lee%
\ \BBA {} Hershberger%
}{%
Lee%
\ \BBA {} Hershberger%
}{%
{\protect \APACyear {1990}}%
}]{%
LeeHershberger1990}
\APACinsertmetastar {%
LeeHershberger1990}%
\begin{APACrefauthors}%
Lee, S.%
\BCBT {}\ \BBA {} Hershberger, S.%
\end{APACrefauthors}%
\unskip\
\newblock
\APACrefYearMonthDay{1990}{}{}.
\newblock
{\BBOQ}\APACrefatitle {A Simple Rule for Generating Equivalent Models in
  Covariance Structure Modeling} {A simple rule for generating equivalent
  models in covariance structure modeling}.{\BBCQ}
\newblock
\APACjournalVolNumPages{Multivariate Behavioral Research}{25}{3}{313--334}.
\PrintBackRefs{\CurrentBib}

\bibitem [\protect \citeauthoryear {%
Luo%
, Moffa%
\BCBL {}\ \BBA {} Kuipers%
}{%
Luo%
\ \protect \BOthers {.}}{%
{\protect \APACyear {2021}}%
}]{%
LuoMoffaKuipers2021}
\APACinsertmetastar {%
LuoMoffaKuipers2021}%
\begin{APACrefauthors}%
Luo, X\BPBI G.%
, Moffa, G.%
\BCBL {}\ \BBA {} Kuipers, J.%
\end{APACrefauthors}%
\unskip\
\newblock
\APACrefYearMonthDay{2021}{}{}.
\newblock
{\BBOQ}\APACrefatitle {Learning {Bayesian} Networks from Ordinal Data}
  {Learning {Bayesian} networks from ordinal data}.{\BBCQ}
\newblock
\APACjournalVolNumPages{Journal of Machine Learning Research}{22}{266}{1--44}.
\PrintBackRefs{\CurrentBib}

\bibitem [\protect \citeauthoryear {%
Lyu%
, Wang%
\BCBL {}\ \BBA {} Xu%
}{%
Lyu%
\ \protect \BOthers {.}}{%
{\protect \APACyear {2025}}%
}]{%
LyuWangXu2025}
\APACinsertmetastar {%
LyuWangXu2025}%
\begin{APACrefauthors}%
Lyu, W.%
, Wang, C.%
\BCBL {}\ \BBA {} Xu, G.%
\end{APACrefauthors}%
\unskip\
\newblock
\APACrefYearMonthDay{2025}{}{}.
\newblock
{\BBOQ}\APACrefatitle {Detecting Differential Item Functioning across Multiple
  Groups Using Group Pairwise Penalty} {Detecting differential item functioning
  across multiple groups using group pairwise penalty}.{\BBCQ}
\newblock
\APACjournalVolNumPages{Psychometrika}{90}{5}{1594--1621}.
\PrintBackRefs{\CurrentBib}

\bibitem [\protect \citeauthoryear {%
MacCallum%
}{%
MacCallum%
}{%
{\protect \APACyear {1986}}%
}]{%
MacCallum1986}
\APACinsertmetastar {%
MacCallum1986}%
\begin{APACrefauthors}%
MacCallum, R\BPBI C.%
\end{APACrefauthors}%
\unskip\
\newblock
\APACrefYearMonthDay{1986}{}{}.
\newblock
{\BBOQ}\APACrefatitle {Specification Searches in Covariance Structure Modeling}
  {Specification searches in covariance structure modeling}.{\BBCQ}
\newblock
\APACjournalVolNumPages{Psychological Bulletin}{100}{1}{107--120}.
\PrintBackRefs{\CurrentBib}

\bibitem [\protect \citeauthoryear {%
MacCallum%
, Wegener%
, Uchino%
\BCBL {}\ \BBA {} Fabrigar%
}{%
MacCallum%
\ \protect \BOthers {.}}{%
{\protect \APACyear {1993}}%
}]{%
maccallum1993problem}
\APACinsertmetastar {%
maccallum1993problem}%
\begin{APACrefauthors}%
MacCallum, R\BPBI C.%
, Wegener, D\BPBI T.%
, Uchino, B\BPBI N.%
\BCBL {}\ \BBA {} Fabrigar, L\BPBI R.%
\end{APACrefauthors}%
\unskip\
\newblock
\APACrefYearMonthDay{1993}{}{}.
\newblock
{\BBOQ}\APACrefatitle {The Problem of Equivalent Models in Applications of
  Covariance Structure Analysis} {The problem of equivalent models in
  applications of covariance structure analysis}.{\BBCQ}
\newblock
\APACjournalVolNumPages{Psychological Bulletin}{114}{1}{185--199}.
\PrintBackRefs{\CurrentBib}

\bibitem [\protect \citeauthoryear {%
Marsh%
\ \BBA {} Hau%
}{%
Marsh%
\ \BBA {} Hau%
}{%
{\protect \APACyear {2007}}%
}]{%
MarshHau2007}
\APACinsertmetastar {%
MarshHau2007}%
\begin{APACrefauthors}%
Marsh, H\BPBI W.%
\BCBT {}\ \BBA {} Hau, K\BHBI T.%
\end{APACrefauthors}%
\unskip\
\newblock
\APACrefYearMonthDay{2007}{}{}.
\newblock
{\BBOQ}\APACrefatitle {Applications of Latent-Variable Models in Educational
  Psychology: The Need for Methodological-Substantive Synergies} {Applications
  of latent-variable models in educational psychology: The need for
  methodological-substantive synergies}.{\BBCQ}
\newblock
\APACjournalVolNumPages{Contemporary Educational Psychology}{32}{1}{151--170}.
\PrintBackRefs{\CurrentBib}

\bibitem [\protect \citeauthoryear {%
Marsh%
, Morin%
, Parker%
\BCBL {}\ \BBA {} Kaur%
}{%
Marsh%
\ \protect \BOthers {.}}{%
{\protect \APACyear {2014}}%
}]{%
MarshMorinParkerKaur2014}
\APACinsertmetastar {%
MarshMorinParkerKaur2014}%
\begin{APACrefauthors}%
Marsh, H\BPBI W.%
, Morin, A\BPBI J\BPBI S.%
, Parker, P\BPBI D.%
\BCBL {}\ \BBA {} Kaur, G.%
\end{APACrefauthors}%
\unskip\
\newblock
\APACrefYearMonthDay{2014}{}{}.
\newblock
{\BBOQ}\APACrefatitle {Exploratory Structural Equation Modeling: An Integration
  of the Best Features of Exploratory and Confirmatory Factor Analysis}
  {Exploratory structural equation modeling: An integration of the best
  features of exploratory and confirmatory factor analysis}.{\BBCQ}
\newblock
\APACjournalVolNumPages{Annual Review of Clinical Psychology}{10}{}{85--110}.
\PrintBackRefs{\CurrentBib}

\bibitem [\protect \citeauthoryear {%
Marsman%
\ \protect \BOthers {.}}{%
Marsman%
\ \protect \BOthers {.}}{%
{\protect \APACyear {2018}}%
}]{%
MarsmanEtAl2018}
\APACinsertmetastar {%
MarsmanEtAl2018}%
\begin{APACrefauthors}%
Marsman, M.%
, Borsboom, D.%
, Kruis, J.%
, Epskamp, S.%
, van Bork, R.%
, Waldorp, L\BPBI J.%
\BDBL {}Maris, G\BPBI K\BPBI J.%
\end{APACrefauthors}%
\unskip\
\newblock
\APACrefYearMonthDay{2018}{}{}.
\newblock
{\BBOQ}\APACrefatitle {An Introduction to Network Psychometrics: Relating
  {Ising} Network Models to Item Response Theory Models} {An introduction to
  network psychometrics: Relating {Ising} network models to item response
  theory models}.{\BBCQ}
\newblock
\APACjournalVolNumPages{Multivariate Behavioral Research}{53}{1}{15--35}.
\PrintBackRefs{\CurrentBib}

\bibitem [\protect \citeauthoryear {%
Marsman%
, Huth%
, Waldorp%
\BCBL {}\ \BBA {} Ntzoufras%
}{%
Marsman%
\ \protect \BOthers {.}}{%
{\protect \APACyear {2022}}%
}]{%
MarsmanHuthWaldorpNtzoufras2022}
\APACinsertmetastar {%
MarsmanHuthWaldorpNtzoufras2022}%
\begin{APACrefauthors}%
Marsman, M.%
, Huth, K.%
, Waldorp, L\BPBI J.%
\BCBL {}\ \BBA {} Ntzoufras, I.%
\end{APACrefauthors}%
\unskip\
\newblock
\APACrefYearMonthDay{2022}{}{}.
\newblock
{\BBOQ}\APACrefatitle {Objective {B}ayesian Edge Screening and Structure
  Selection for {I}sing Networks} {Objective {B}ayesian edge screening and
  structure selection for {I}sing networks}.{\BBCQ}
\newblock
\APACjournalVolNumPages{Psychometrika}{87}{1}{47--82}.
\PrintBackRefs{\CurrentBib}

\bibitem [\protect \citeauthoryear {%
Marsman%
\ \BBA {} Rhemtulla%
}{%
Marsman%
\ \BBA {} Rhemtulla%
}{%
{\protect \APACyear {2022}}%
}]{%
MarsmanRhemtulla2022}
\APACinsertmetastar {%
MarsmanRhemtulla2022}%
\begin{APACrefauthors}%
Marsman, M.%
\BCBT {}\ \BBA {} Rhemtulla, M.%
\end{APACrefauthors}%
\unskip\
\newblock
\APACrefYearMonthDay{2022}{}{}.
\newblock
{\BBOQ}\APACrefatitle {Guest Editors' Introduction to the Special Issue
  ``{N}etwork Psychometrics in Action'': Methodological Innovations Inspired by
  Empirical Problems} {Guest editors' introduction to the special issue
  ``{N}etwork psychometrics in action'': Methodological innovations inspired by
  empirical problems}.{\BBCQ}
\newblock
\APACjournalVolNumPages{Psychometrika}{87}{1}{1--11}.
\PrintBackRefs{\CurrentBib}

\bibitem [\protect \citeauthoryear {%
Marsman%
, van~den Bergh%
\BCBL {}\ \BBA {} Haslbeck%
}{%
Marsman%
\ \protect \BOthers {.}}{%
{\protect \APACyear {2025}}%
}]{%
MarsmanVanDenBerghHaslbeck2025}
\APACinsertmetastar {%
MarsmanVanDenBerghHaslbeck2025}%
\begin{APACrefauthors}%
Marsman, M.%
, van~den Bergh, D.%
\BCBL {}\ \BBA {} Haslbeck, J\BPBI M\BPBI B.%
\end{APACrefauthors}%
\unskip\
\newblock
\APACrefYearMonthDay{2025}{}{}.
\newblock
{\BBOQ}\APACrefatitle {Bayesian Analysis of the Ordinal {M}arkov Random Field}
  {Bayesian analysis of the ordinal {M}arkov random field}.{\BBCQ}
\newblock
\APACjournalVolNumPages{Psychometrika}{90}{1}{146--182}.
\PrintBackRefs{\CurrentBib}

\bibitem [\protect \citeauthoryear {%
Meek%
}{%
Meek%
}{%
{\protect \APACyear {1995}}%
{\protect \APACexlab {{\protect \BCnt {1}}}}}]{%
meek1995causal}
\APACinsertmetastar {%
meek1995causal}%
\begin{APACrefauthors}%
Meek, C.%
\end{APACrefauthors}%
\unskip\
\newblock
\APACrefYearMonthDay{1995{\protect \BCnt {1}}}{}{}.
\newblock
{\BBOQ}\APACrefatitle {Causal Inference and Causal Explanation with Background
  Knowledge} {Causal inference and causal explanation with background
  knowledge}.{\BBCQ}
\newblock
\BIn{} \APACrefbtitle {Proceedings of the Eleventh Conference on Uncertainty in
  Artificial Intelligence} {Proceedings of the eleventh conference on
  uncertainty in artificial intelligence}\ (\BPGS\ 403--410).
\newblock
\APACaddressPublisher{}{Morgan Kaufmann}.
\PrintBackRefs{\CurrentBib}

\bibitem [\protect \citeauthoryear {%
Meek%
}{%
Meek%
}{%
{\protect \APACyear {1995}}%
{\protect \APACexlab {{\protect \BCnt {2}}}}}]{%
meek1995strong}
\APACinsertmetastar {%
meek1995strong}%
\begin{APACrefauthors}%
Meek, C.%
\end{APACrefauthors}%
\unskip\
\newblock
\APACrefYearMonthDay{1995{\protect \BCnt {2}}}{}{}.
\newblock
{\BBOQ}\APACrefatitle {Strong Completeness and Faithfulness in {B}ayesian
  Networks} {Strong completeness and faithfulness in {B}ayesian
  networks}.{\BBCQ}
\newblock
\BIn{} \APACrefbtitle {Proceedings of the Eleventh Conference on Uncertainty in
  Artificial Intelligence} {Proceedings of the eleventh conference on
  uncertainty in artificial intelligence}\ (\BPGS\ 411--418).
\newblock
\APACaddressPublisher{}{Morgan Kaufmann}.
\PrintBackRefs{\CurrentBib}

\bibitem [\protect \citeauthoryear {%
Morin%
, Marsh%
\BCBL {}\ \BBA {} Nagengast%
}{%
Morin%
\ \protect \BOthers {.}}{%
{\protect \APACyear {2013}}%
}]{%
Morinetal:13}
\APACinsertmetastar {%
Morinetal:13}%
\begin{APACrefauthors}%
Morin, A\BPBI J\BPBI S.%
, Marsh, H\BPBI W.%
\BCBL {}\ \BBA {} Nagengast, B.%
\end{APACrefauthors}%
\unskip\
\newblock
\APACrefYearMonthDay{2013}{}{}.
\newblock
{\BBOQ}\APACrefatitle {Exploratory Structural Equation Modeling} {Exploratory
  structural equation modeling}.{\BBCQ}
\newblock
\BIn{} G\BPBI R.~Hancock\ \BBA {} R\BPBI O.~Mueller\ (\BEDS), \APACrefbtitle
  {Structural Equation Modeling: A Second Course} {Structural equation
  modeling: A second course}\ (\PrintOrdinal{2nd}\ \BEd, \BPGS\ 395--436).
\newblock
\APACaddressPublisher{Charlotte, NC}{Information Age Publishing}.
\PrintBackRefs{\CurrentBib}

\bibitem [\protect \citeauthoryear {%
Muth{\'e}n%
\ \BBA {} Asparouhov%
}{%
Muth{\'e}n%
\ \BBA {} Asparouhov%
}{%
{\protect \APACyear {2012}}%
}]{%
MuthenAsparouhov2012}
\APACinsertmetastar {%
MuthenAsparouhov2012}%
\begin{APACrefauthors}%
Muth{\'e}n, B.%
\BCBT {}\ \BBA {} Asparouhov, T.%
\end{APACrefauthors}%
\unskip\
\newblock
\APACrefYearMonthDay{2012}{}{}.
\newblock
{\BBOQ}\APACrefatitle {Bayesian Structural Equation Modeling: A More Flexible
  Representation of Substantive Theory} {Bayesian structural equation modeling:
  A more flexible representation of substantive theory}.{\BBCQ}
\newblock
\APACjournalVolNumPages{Psychological Methods}{17}{3}{313--335}.
\PrintBackRefs{\CurrentBib}

\bibitem [\protect \citeauthoryear {%
Ni%
, Chen%
\BCBL {}\ \BBA {} Wang%
}{%
Ni%
\ \protect \BOthers {.}}{%
{\protect \APACyear {2025}}%
}]{%
ni2025causal}
\APACinsertmetastar {%
ni2025causal}%
\begin{APACrefauthors}%
Ni, Y.%
, Chen, S.%
\BCBL {}\ \BBA {} Wang, Z.%
\end{APACrefauthors}%
\unskip\
\newblock
\APACrefYearMonthDay{2025}{}{}.
\newblock
{\BBOQ}\APACrefatitle {Causal Structural Modeling of Survey Questionnaires via
  a Bootstrapped Ordinal {B}ayesian Network Approach} {Causal structural
  modeling of survey questionnaires via a bootstrapped ordinal {B}ayesian
  network approach}.{\BBCQ}
\newblock
\APACjournalVolNumPages{Psychometrika}{90}{1}{229--250}.
\PrintBackRefs{\CurrentBib}

\bibitem [\protect \citeauthoryear {%
Nocedal%
\ \BBA {} Wright%
}{%
Nocedal%
\ \BBA {} Wright%
}{%
{\protect \APACyear {2006}}%
}]{%
nocedal2006numerical}
\APACinsertmetastar {%
nocedal2006numerical}%
\begin{APACrefauthors}%
Nocedal, J.%
\BCBT {}\ \BBA {} Wright, S\BPBI J.%
\end{APACrefauthors}%
\unskip\
\newblock
\APACrefYear{2006}.
\newblock
\APACrefbtitle {Numerical Optimization} {Numerical optimization}\
  (\PrintOrdinal{2nd}\ \BEd).
\newblock
\APACaddressPublisher{New York}{Springer}.
\PrintBackRefs{\CurrentBib}

\bibitem [\protect \citeauthoryear {%
Park%
, Waldorp%
\BCBL {}\ \BBA {} Ryan%
}{%
Park%
\ \protect \BOthers {.}}{%
{\protect \APACyear {2024}}%
}]{%
ParkWaldorpRyan2024}
\APACinsertmetastar {%
ParkWaldorpRyan2024}%
\begin{APACrefauthors}%
Park, K.%
, Waldorp, L\BPBI J.%
\BCBL {}\ \BBA {} Ryan, O.%
\end{APACrefauthors}%
\unskip\
\newblock
\APACrefYearMonthDay{2024}{}{}.
\newblock
{\BBOQ}\APACrefatitle {Discovering Cyclic Causal Models in Psychological
  Research} {Discovering cyclic causal models in psychological
  research}.{\BBCQ}
\newblock
\APACjournalVolNumPages{advances.in/psychology}{2}{1}{e72425}.
\newblock
\begin{APACrefDOI} \doi{10.56296/aip00012} \end{APACrefDOI}
\PrintBackRefs{\CurrentBib}

\bibitem [\protect \citeauthoryear {%
Pearl%
}{%
Pearl%
}{%
{\protect \APACyear {1988}}%
}]{%
Pearl1988}
\APACinsertmetastar {%
Pearl1988}%
\begin{APACrefauthors}%
Pearl, J.%
\end{APACrefauthors}%
\unskip\
\newblock
\APACrefYear{1988}.
\newblock
\APACrefbtitle {Probabilistic Reasoning in Intelligent Systems: Networks of
  Plausible Inference} {Probabilistic reasoning in intelligent systems:
  Networks of plausible inference}.
\newblock
\APACaddressPublisher{San Mateo, CA}{Morgan Kaufmann}.
\PrintBackRefs{\CurrentBib}

\bibitem [\protect \citeauthoryear {%
Pearl%
}{%
Pearl%
}{%
{\protect \APACyear {2009}}%
}]{%
pearl2009causality}
\APACinsertmetastar {%
pearl2009causality}%
\begin{APACrefauthors}%
Pearl, J.%
\end{APACrefauthors}%
\unskip\
\newblock
\APACrefYear{2009}.
\newblock
\APACrefbtitle {Causality: Models, Reasoning, and Inference} {Causality:
  Models, reasoning, and inference}\ (\PrintOrdinal{2nd}\ \BEd).
\newblock
\APACaddressPublisher{Cambridge}{Cambridge University Press}.
\PrintBackRefs{\CurrentBib}

\bibitem [\protect \citeauthoryear {%
Ryan%
, Bringmann%
\BCBL {}\ \BBA {} Schuurman%
}{%
Ryan%
\ \protect \BOthers {.}}{%
{\protect \APACyear {2022}}%
}]{%
RyanBringmannSchuurman2022}
\APACinsertmetastar {%
RyanBringmannSchuurman2022}%
\begin{APACrefauthors}%
Ryan, O.%
, Bringmann, L\BPBI F.%
\BCBL {}\ \BBA {} Schuurman, N\BPBI K.%
\end{APACrefauthors}%
\unskip\
\newblock
\APACrefYearMonthDay{2022}{}{}.
\newblock
{\BBOQ}\APACrefatitle {The Challenge of Generating Causal Hypotheses Using
  Network Models} {The challenge of generating causal hypotheses using network
  models}.{\BBCQ}
\newblock
\APACjournalVolNumPages{Structural Equation Modeling: A Multidisciplinary
  Journal}{29}{6}{953--970}.
\PrintBackRefs{\CurrentBib}

\bibitem [\protect \citeauthoryear {%
Shen%
, Pan%
\BCBL {}\ \BBA {} Zhu%
}{%
Shen%
\ \protect \BOthers {.}}{%
{\protect \APACyear {2012}}%
}]{%
shen2012likelihood}
\APACinsertmetastar {%
shen2012likelihood}%
\begin{APACrefauthors}%
Shen, X.%
, Pan, W.%
\BCBL {}\ \BBA {} Zhu, Y.%
\end{APACrefauthors}%
\unskip\
\newblock
\APACrefYearMonthDay{2012}{}{}.
\newblock
{\BBOQ}\APACrefatitle {Likelihood-Based Selection and Sharp Parameter
  Estimation} {Likelihood-based selection and sharp parameter
  estimation}.{\BBCQ}
\newblock
\APACjournalVolNumPages{Journal of the American Statistical
  Association}{107}{497}{223--232}.
\PrintBackRefs{\CurrentBib}

\bibitem [\protect \citeauthoryear {%
Shen%
, Pan%
, Zhu%
\BCBL {}\ \BBA {} Zhou%
}{%
Shen%
\ \protect \BOthers {.}}{%
{\protect \APACyear {2013}}%
}]{%
shen2013constrained}
\APACinsertmetastar {%
shen2013constrained}%
\begin{APACrefauthors}%
Shen, X.%
, Pan, W.%
, Zhu, Y.%
\BCBL {}\ \BBA {} Zhou, H.%
\end{APACrefauthors}%
\unskip\
\newblock
\APACrefYearMonthDay{2013}{}{}.
\newblock
{\BBOQ}\APACrefatitle {On Constrained and Regularized High-Dimensional
  Regression} {On constrained and regularized high-dimensional
  regression}.{\BBCQ}
\newblock
\APACjournalVolNumPages{Annals of the Institute of Statistical
  Mathematics}{65}{5}{807--832}.
\PrintBackRefs{\CurrentBib}

\bibitem [\protect \citeauthoryear {%
Silva%
, Scheines%
, Glymour%
\BCBL {}\ \BBA {} Spirtes%
}{%
Silva%
\ \protect \BOthers {.}}{%
{\protect \APACyear {2006}}%
}]{%
SilvaScheinesGlymourSpirtes2006}
\APACinsertmetastar {%
SilvaScheinesGlymourSpirtes2006}%
\begin{APACrefauthors}%
Silva, R.%
, Scheines, R.%
, Glymour, C.%
\BCBL {}\ \BBA {} Spirtes, P.%
\end{APACrefauthors}%
\unskip\
\newblock
\APACrefYearMonthDay{2006}{}{}.
\newblock
{\BBOQ}\APACrefatitle {Learning the Structure of Linear Latent Variable Models}
  {Learning the structure of linear latent variable models}.{\BBCQ}
\newblock
\APACjournalVolNumPages{Journal of Machine Learning Research}{7}{}{191--246}.
\PrintBackRefs{\CurrentBib}

\bibitem [\protect \citeauthoryear {%
Skrondal%
\ \BBA {} Laake%
}{%
Skrondal%
\ \BBA {} Laake%
}{%
{\protect \APACyear {2001}}%
}]{%
SkrondalLaake2001}
\APACinsertmetastar {%
SkrondalLaake2001}%
\begin{APACrefauthors}%
Skrondal, A.%
\BCBT {}\ \BBA {} Laake, P.%
\end{APACrefauthors}%
\unskip\
\newblock
\APACrefYearMonthDay{2001}{}{}.
\newblock
{\BBOQ}\APACrefatitle {Regression Among Factor Scores} {Regression among factor
  scores}.{\BBCQ}
\newblock
\APACjournalVolNumPages{Psychometrika}{66}{4}{563--575}.
\PrintBackRefs{\CurrentBib}

\bibitem [\protect \citeauthoryear {%
S{\"o}rbom%
}{%
S{\"o}rbom%
}{%
{\protect \APACyear {1989}}%
}]{%
Sorbom1989}
\APACinsertmetastar {%
Sorbom1989}%
\begin{APACrefauthors}%
S{\"o}rbom, D.%
\end{APACrefauthors}%
\unskip\
\newblock
\APACrefYearMonthDay{1989}{}{}.
\newblock
{\BBOQ}\APACrefatitle {Model Modification} {Model modification}.{\BBCQ}
\newblock
\APACjournalVolNumPages{Psychometrika}{54}{3}{371--384}.
\PrintBackRefs{\CurrentBib}

\bibitem [\protect \citeauthoryear {%
Spirtes%
, Glymour%
\BCBL {}\ \BBA {} Scheines%
}{%
Spirtes%
\ \protect \BOthers {.}}{%
{\protect \APACyear {2000}}%
}]{%
SpirtesGlymourScheines2000}
\APACinsertmetastar {%
SpirtesGlymourScheines2000}%
\begin{APACrefauthors}%
Spirtes, P.%
, Glymour, C\BPBI N.%
\BCBL {}\ \BBA {} Scheines, R.%
\end{APACrefauthors}%
\unskip\
\newblock
\APACrefYear{2000}.
\newblock
\APACrefbtitle {Causation, Prediction, and Search} {Causation, prediction, and
  search}\ (\PrintOrdinal{2nd}\ \BEd).
\newblock
\APACaddressPublisher{Cambridge, MA}{The MIT Press}.
\PrintBackRefs{\CurrentBib}

\bibitem [\protect \citeauthoryear {%
Stelzl%
}{%
Stelzl%
}{%
{\protect \APACyear {1986}}%
}]{%
Stelzl1986}
\APACinsertmetastar {%
Stelzl1986}%
\begin{APACrefauthors}%
Stelzl, I.%
\end{APACrefauthors}%
\unskip\
\newblock
\APACrefYearMonthDay{1986}{}{}.
\newblock
{\BBOQ}\APACrefatitle {Changing a Causal Hypothesis Without Changing the Fit:
  Some Rules for Generating Equivalent Path Models} {Changing a causal
  hypothesis without changing the fit: Some rules for generating equivalent
  path models}.{\BBCQ}
\newblock
\APACjournalVolNumPages{Multivariate Behavioral Research}{21}{3}{309--331}.
\PrintBackRefs{\CurrentBib}

\bibitem [\protect \citeauthoryear {%
Sunshine%
\ \BBA {} Tyler%
}{%
Sunshine%
\ \BBA {} Tyler%
}{%
{\protect \APACyear {2003}}%
}]{%
sunshine2003role}
\APACinsertmetastar {%
sunshine2003role}%
\begin{APACrefauthors}%
Sunshine, J.%
\BCBT {}\ \BBA {} Tyler, T\BPBI R.%
\end{APACrefauthors}%
\unskip\
\newblock
\APACrefYearMonthDay{2003}{}{}.
\newblock
{\BBOQ}\APACrefatitle {The Role of Procedural Justice and Legitimacy in Shaping
  Public Support for Policing} {The role of procedural justice and legitimacy
  in shaping public support for policing}.{\BBCQ}
\newblock
\APACjournalVolNumPages{Law \& Society Review}{37}{3}{513--547}.
\PrintBackRefs{\CurrentBib}

\bibitem [\protect \citeauthoryear {%
Tyler%
}{%
Tyler%
}{%
{\protect \APACyear {2006}}%
}]{%
tyler2006why}
\APACinsertmetastar {%
tyler2006why}%
\begin{APACrefauthors}%
Tyler, T\BPBI R.%
\end{APACrefauthors}%
\unskip\
\newblock
\APACrefYear{2006}.
\newblock
\APACrefbtitle {Why People Obey the Law} {Why people obey the law}.
\newblock
\APACaddressPublisher{Princeton, NJ}{Princeton University Press}.
\PrintBackRefs{\CurrentBib}

\bibitem [\protect \citeauthoryear {%
van Bork%
\ \protect \BOthers {.}}{%
van Bork%
\ \protect \BOthers {.}}{%
{\protect \APACyear {2021}}%
}]{%
VanBorkEtAl2021}
\APACinsertmetastar {%
VanBorkEtAl2021}%
\begin{APACrefauthors}%
van Bork, R.%
, Rhemtulla, M.%
, Waldorp, L\BPBI J.%
, Kruis, J.%
, Rezvanifar, S.%
\BCBL {}\ \BBA {} Borsboom, D.%
\end{APACrefauthors}%
\unskip\
\newblock
\APACrefYearMonthDay{2021}{}{}.
\newblock
{\BBOQ}\APACrefatitle {Latent Variable Models and Networks: Statistical
  Equivalence and Testability} {Latent variable models and networks:
  Statistical equivalence and testability}.{\BBCQ}
\newblock
\APACjournalVolNumPages{Multivariate Behavioral Research}{56}{2}{175--198}.
\PrintBackRefs{\CurrentBib}

\bibitem [\protect \citeauthoryear {%
van~de Geer%
\ \BBA {} B{\"u}hlmann%
}{%
van~de Geer%
\ \BBA {} B{\"u}hlmann%
}{%
{\protect \APACyear {2013}}%
}]{%
vandegeer2013l0}
\APACinsertmetastar {%
vandegeer2013l0}%
\begin{APACrefauthors}%
van~de Geer, S.%
\BCBT {}\ \BBA {} B{\"u}hlmann, P.%
\end{APACrefauthors}%
\unskip\
\newblock
\APACrefYearMonthDay{2013}{}{}.
\newblock
{\BBOQ}\APACrefatitle {$\ell_0$-Penalized Maximum Likelihood for Sparse
  Directed Acyclic Graphs} {$\ell_0$-penalized maximum likelihood for sparse
  directed acyclic graphs}.{\BBCQ}
\newblock
\APACjournalVolNumPages{The Annals of Statistics}{41}{2}{536--567}.
\PrintBackRefs{\CurrentBib}

\bibitem [\protect \citeauthoryear {%
Verma%
\ \BBA {} Pearl%
}{%
Verma%
\ \BBA {} Pearl%
}{%
{\protect \APACyear {1990}}%
}]{%
verma1990equivalence}
\APACinsertmetastar {%
verma1990equivalence}%
\begin{APACrefauthors}%
Verma, T\BPBI S.%
\BCBT {}\ \BBA {} Pearl, J.%
\end{APACrefauthors}%
\unskip\
\newblock
\APACrefYearMonthDay{1990}{}{}.
\newblock
{\BBOQ}\APACrefatitle {Equivalence and Synthesis of Causal Models} {Equivalence
  and synthesis of causal models}.{\BBCQ}
\newblock
\BIn{} \APACrefbtitle {Proceedings of the Sixth Conference on Uncertainty in
  Artificial Intelligence} {Proceedings of the sixth conference on uncertainty
  in artificial intelligence}\ (\BPGS\ 220--227).
\PrintBackRefs{\CurrentBib}

\bibitem [\protect \citeauthoryear {%
Waldorp%
\ \BBA {} Marsman%
}{%
Waldorp%
\ \BBA {} Marsman%
}{%
{\protect \APACyear {2022}}%
}]{%
WaldorpMarsman2022}
\APACinsertmetastar {%
WaldorpMarsman2022}%
\begin{APACrefauthors}%
Waldorp, L.%
\BCBT {}\ \BBA {} Marsman, M.%
\end{APACrefauthors}%
\unskip\
\newblock
\APACrefYearMonthDay{2022}{}{}.
\newblock
{\BBOQ}\APACrefatitle {Relations Between Networks, Regression, Partial
  Correlation, and the Latent Variable Model} {Relations between networks,
  regression, partial correlation, and the latent variable model}.{\BBCQ}
\newblock
\APACjournalVolNumPages{Multivariate Behavioral Research}{57}{6}{994--1006}.
\PrintBackRefs{\CurrentBib}

\bibitem [\protect \citeauthoryear {%
Wermuth%
}{%
Wermuth%
}{%
{\protect \APACyear {1980}}%
}]{%
Wermuth1980}
\APACinsertmetastar {%
Wermuth1980}%
\begin{APACrefauthors}%
Wermuth, N.%
\end{APACrefauthors}%
\unskip\
\newblock
\APACrefYearMonthDay{1980}{}{}.
\newblock
{\BBOQ}\APACrefatitle {Linear Recursive Equations, Covariance Selection, and
  Path Analysis} {Linear recursive equations, covariance selection, and path
  analysis}.{\BBCQ}
\newblock
\APACjournalVolNumPages{Journal of the American Statistical
  Association}{75}{372}{963--972}.
\PrintBackRefs{\CurrentBib}

\bibitem [\protect \citeauthoryear {%
Xie%
, Chen%
, von Davier%
\BCBL {}\ \BBA {} Weng%
}{%
Xie%
\ \protect \BOthers {.}}{%
{\protect \APACyear {2024}}%
}]{%
XieChenVonDavierWeng2024}
\APACinsertmetastar {%
XieChenVonDavierWeng2024}%
\begin{APACrefauthors}%
Xie, Z.%
, Chen, Y.%
, von Davier, M.%
\BCBL {}\ \BBA {} Weng, H.%
\end{APACrefauthors}%
\unskip\
\newblock
\APACrefYearMonthDay{2024}{}{}.
\newblock
{\BBOQ}\APACrefatitle {Variable Selection in Latent Variable Models via
  Knockoffs: An Application to International Large-Scale Assessment in
  Education} {Variable selection in latent variable models via knockoffs: An
  application to international large-scale assessment in education}.{\BBCQ}
\newblock
\APACjournalVolNumPages{Journal of the Royal Statistical Society Series A:
  Statistics in Society}{187}{3}{723--747}.
\PrintBackRefs{\CurrentBib}

\bibitem [\protect \citeauthoryear {%
Yuan%
, Shen%
, Pan%
\BCBL {}\ \BBA {} Wang%
}{%
Yuan%
\ \protect \BOthers {.}}{%
{\protect \APACyear {2019}}%
}]{%
yuan2019constrained}
\APACinsertmetastar {%
yuan2019constrained}%
\begin{APACrefauthors}%
Yuan, Y.%
, Shen, X.%
, Pan, W.%
\BCBL {}\ \BBA {} Wang, Z.%
\end{APACrefauthors}%
\unskip\
\newblock
\APACrefYearMonthDay{2019}{}{}.
\newblock
{\BBOQ}\APACrefatitle {Constrained Likelihood for Reconstructing a Directed
  Acyclic {G}aussian Graph} {Constrained likelihood for reconstructing a
  directed acyclic {G}aussian graph}.{\BBCQ}
\newblock
\APACjournalVolNumPages{Biometrika}{106}{1}{109--125}.
\PrintBackRefs{\CurrentBib}

\bibitem [\protect \citeauthoryear {%
Zhang%
, Chen%
\BCBL {}\ \BBA {} Moustaki%
}{%
Zhang%
\ \protect \BOthers {.}}{%
{\protect \APACyear {2026}}%
}]{%
ZhangChenMoustaki2026}
\APACinsertmetastar {%
ZhangChenMoustaki2026}%
\begin{APACrefauthors}%
Zhang, S.%
, Chen, Y.%
\BCBL {}\ \BBA {} Moustaki, I.%
\end{APACrefauthors}%
\unskip\
\newblock
\APACrefYearMonthDay{2026}{}{}.
\newblock
{\BBOQ}\APACrefatitle {{DAG-CLIP}: A {DAG} Learning Framework in the Presence
  of Latent Variables} {{DAG-CLIP}: A {DAG} learning framework in the presence
  of latent variables}.{\BBCQ}
\newblock
\APACjournalVolNumPages{arXiv preprint arXiv:2610.06086}{}{}{}.
\PrintBackRefs{\CurrentBib}

\bibitem [\protect \citeauthoryear {%
Zhang%
, Wang%
\BCBL {}\ \BBA {} Gu%
}{%
Zhang%
\ \protect \BOthers {.}}{%
{\protect \APACyear {2026}}%
}]{%
zhang2026discrete}
\APACinsertmetastar {%
zhang2026discrete}%
\begin{APACrefauthors}%
Zhang, W.%
, Wang, Y.%
\BCBL {}\ \BBA {} Gu, Y.%
\end{APACrefauthors}%
\unskip\
\newblock
\APACrefYearMonthDay{2026}{}{}.
\newblock
{\BBOQ}\APACrefatitle {Discrete Causal Representation Learning} {Discrete
  causal representation learning}.{\BBCQ}
\newblock
\APACjournalVolNumPages{arXiv preprint arXiv:2603.25017}{}{}{}.
\PrintBackRefs{\CurrentBib}

\bibitem [\protect \citeauthoryear {%
Zheng%
, Aragam%
, Ravikumar%
\BCBL {}\ \BBA {} Xing%
}{%
Zheng%
\ \protect \BOthers {.}}{%
{\protect \APACyear {2018}}%
}]{%
ZhengEtAl2018}
\APACinsertmetastar {%
ZhengEtAl2018}%
\begin{APACrefauthors}%
Zheng, X.%
, Aragam, B.%
, Ravikumar, P\BPBI K.%
\BCBL {}\ \BBA {} Xing, E\BPBI P.%
\end{APACrefauthors}%
\unskip\
\newblock
\APACrefYearMonthDay{2018}{}{}.
\newblock
{\BBOQ}\APACrefatitle {{DAGs with NO TEARS}: Continuous Optimization for
  Structure Learning} {{DAGs with NO TEARS}: Continuous optimization for
  structure learning}.{\BBCQ}
\newblock
\BIn{} \APACrefbtitle {Advances in Neural Information Processing Systems}
  {Advances in neural information processing systems}\ (\BVOL~31, \BPGS\
  9472--9483).
\PrintBackRefs{\CurrentBib}

\end{thebibliography}

\end{document}